\documentclass[twocolumn,showpacs,preprintnumbers,amsmath,amssymb,APSl,prd,nofootinbib,superscriptaddress]{revtex4-1}

\usepackage{bm}
\usepackage{mathrsfs}
\usepackage{xcolor,color,graphicx,graphics}
\usepackage[all]{xy}
\usepackage{epsfig,subfigure}
\usepackage{latexsym,amssymb,amsmath,amsfonts} 
\usepackage[english]{babel} 
\usepackage[OT1]{fontenc}
\usepackage[latin1]{inputenc}
\usepackage{makeidx}
\usepackage{hyperref}
\usepackage{color,graphicx,graphics,wrapfig,epsf}

\definecolor{red}{rgb}{1,0,0}

\def\p{\partial}

\def\+{^\dagger}

\def\<{\leftarrow}
\def\>{\rightarrow}

\def\({\left(}
\def\){\right)}

\def\a{\alpha} \def\b{\beta}  \def\d{\delta} 
\def\m{\mu} \def\n{\nu}   \def\l{\lambda} 
\def\k{\kappa}\def\G{\Gamma}

\def\q{\quad}\def\qq{\qquad}

\def\A{{\cal A}}\def\B{{\cal B}}

\def\W{{\Omega}}
\newcommand{\LL}{{\cal L}}

\newcommand{\bi}{\begin{itemize}} 				\newcommand{\ei}{\end{itemize}}
\newcommand{\benu}{\begin{enumerate}} 		\newcommand{\enu}{\end{enumerate}}
\newcommand{\bd}{\begin{dinglist}{0}}     \newcommand{\ed}{\end{dinglist}}
\newcommand{\bfig}{\begin{figure}[htbp]}  \newcommand{\efig}{\end{figure}}
        			
\newcommand{\bc}{\begin{center}} 				  \newcommand{\ec}{\end{center}}
\newcommand{\be}{\begin{equation}} 				\newcommand{\ee}{\end{equation}}
\newcommand{\bsub}{\begin{subequations}}  \newcommand{\esub}{\end{subequations}}
\newcommand{\ben}{\begin{eqnarray}} 			\newcommand{\een}{\end{eqnarray}}
\newcommand{\ba}[1]{\begin{array}{#1}} 		\newcommand{\ea}{\end{array}}
\newcommand{\bea}{\begin{equation}\begin{array}{rcl}}
\newcommand{\eea}{\end{array}\end{equation}}

\begin{document}
\title{A correspondence between modified gravity and General Relativity with scalar fields}

\author{Victor I. Afonso} \email{viafonso@df.ufcg.edu.br}
\affiliation{Unidade Acad\^{e}mica de F\'isica, Universidade Federal de Campina
Grande, 58429-900 Campina Grande, PB, Brazil}
\author{Gonzalo J. Olmo} \email{gonzalo.olmo@uv.es}
\affiliation{Departamento de F\'{i}sica Te\'{o}rica and IFIC, Centro Mixto Universidad de Valencia - CSIC.
Universidad de Valencia, Burjassot-46100, Valencia, Spain}
\affiliation{Departamento de F\'isica, Universidade Federal da
Para\'\i ba, 58051-900 Jo\~ao Pessoa, Para\'\i ba, Brazil}
\author{Emanuele Orazi} \email{orazi.emanuele@gmail.com}
\affiliation{ International Institute of Physics, Federal University of Rio Grande do Norte,
Campus Universit\'ario-Lagoa Nova, Natal-RN 59078-970, Brazil}
\affiliation{Escola de Ciencia e Tecnologia, Universidade Federal do Rio Grande do Norte, Caixa Postal 1524, Natal-RN 59078-970, Brazil}
\author{Diego Rubiera-Garcia} \email{drgarcia@fc.ul.pt}
\affiliation{Instituto de Astrof\'{\i}sica e Ci\^{e}ncias do Espa\c{c}o, Faculdade de
Ci\^encias da Universidade de Lisboa, Edif\'{\i}cio C8, Campo Grande,
P-1749-016 Lisbon, Portugal}

\date{\today}
\begin{abstract}
We describe a novel procedure to map the field equations of nonlinear Ricci-based metric-affine theories of gravity, coupled to scalar matter described by a given Lagrangian, into the field equations of General Relativity coupled to a different scalar field Lagrangian. Our analysis considers examples with a single and $N$ real scalar fields, described either by canonical Lagrangians or by generalized functions of the kinetic and potential terms. In particular, we consider several explicit examples involving $f(R)$ theories and the Eddington-inspired Born-Infeld gravity model, coupled to different scalar field Lagrangians. We show how the nonlinearities of the gravitational sector of these theories can be traded to nonlinearities in the matter fields, and how the procedure allows to find new solutions on both sides of the correspondence.         The potential of this procedure for applications of scalar field models in astrophysiftcal and cosmological scenarios is highlighted.
\end{abstract}

\maketitle

\section{Introduction} \label{sec:II}

In the wake of gravitational wave astronomy after the observation of binary black hole \cite{Abbott:2016blz,Abbott:2017oio} and neutron star mergers \cite{TheLIGOScientific:2017qsa} by the LIGO/Virgo Collaboration, and the future launching of new cosmological probes such as EUCLID \cite{Laureijs:2011gra,Amendola:2016saw}, many of the gravitational extensions of General Relativity (GR) proposed in the literature will be put to experimental test in astrophysical \cite{Barack:2018yly,SeyYag18}, extragalactic  \cite{Collett:2018gpf} and cosmological \cite{Ishak:2018his} contexts, thus going beyond the classical solar system ones \cite{Will:2014kxa}. Indeed, the recent combined gravitational and electromagnetic observations from a neutron star merger have already been able to rule out many of the most popular such extensions, and to put strong constraints upon many others \cite{Lombriser:2015sxa, Lombriser:2016yzn, Baker:2017hug,Sakstein:2017xjx,Creminelli:2017sry} (see the enlightening discussion in \cite{Ezquiaga:2017ekz,EzZuma}). As the pool of observationally viable theories of gravity beyond GR diminishes, there is more than ever a need to rethink the underlying physical and gravitational principles under which such models are formulated, which has triggered the investigation of a number of alternatives \cite{Ferraro:2006jd,Maluf:2013gaa,Bamba:2013ooa,BeltranJimenez:2017tkd,Jarv:2018bgs,BeltranJimenez:2018vdo}. The present work focuses on the formulation of gravitational models where metric and affine connection are independent objects (commonly known as metric-affine or Palatini-formulated theories \cite{Olmo:2011uz}), which has been comparatively much less explored in the literature than their metric cousins, where the affine connection is taken to be given by the Christoffel symbols of the metric \emph{ab initio}.

The analysis of metric-affine extensions of GR has so far been almost exclusively restricted to Lagrangians involving functions of the metric and the Ricci tensor (Ricci-Based Gravities, or RBGs, for short). To our knowledge, the only exceptions are the so-called Lovelock theories \cite{Deruelle,Borunda:2008kf,Charmousis08} and some scalar-tensor models involving explicitly the nonmetricity tensor (covariant derivatives of the metric)  \cite{Burton:1997sj}. The reasons behind this limitation can be found on the difficulties to obtain solutions for the connection equation. In fact, efficient algorithms have only been implemented for RBGs, whereas for theories involving the Riemann tensor and/or other objects, the analysis has simply been limited to verifying the existence of some solutions, not to prove their uniqueness by any means.

In the RBG framework, the fact that (part of) the connection equation can be solved in terms of an auxiliary metric $q_{\mu\nu}$, has allowed to identify the existence of an Einstein frame for  these theories. This Einstein frame is useful as long as it can be used to write the field equations of the corresponding gravity theory in a compact form, namely, in terms of the Einstein tensor of the auxiliary metric on the left-hand-side and everything else on the right-hand-side. The latter is made out of the stress-energy tensor of the matter fields, the spacetime metric $g_{\mu\nu}$, and possibly the auxiliary metric $q_{\mu\nu}$ as well. The fact that $g_{\mu\nu}$ cannot always be explicitly expressed in terms of $q_{\mu\nu}$ and the matter fields is an important drawback, as it forces one to deal with cumbersome equations and rely on the existence of the particular simplifications that may arise in scenarios with specific symmetries. This is the case, for instance, of homogeneous and isotropic cosmological models  \cite{Odintsov:2014yaa,BeltranJimenez:2017uwv}, and of static spherically symmetric spacetimes \cite{Olmo:2011ja,Olmo:2013gqa,Bambi:2015zch}. Any other more sophisticated (or physical) considerations make it impossible in practice to try and solve the field equations. This essential difficulty is precluding further progress on the implementation of astrophysical and cosmological applications of these models. In particular, the application of numerical methods to explore dynamical scenarios such as binary black hole/neutron stars mergers and the generation of gravitational waves appears as a daunting task, requiring the development of specific methods to fit the peculiarities of each particular model.

In a recent article \cite{Afonso:2018bpv} we pointed out that there is a systematic way to avoid the difficulties described above. It turns out that it is possible to establish a correspondence between the space of solutions of an arbitrary RBG coupled to a certain matter source, and the space of solutions of GR coupled to that same source but with a modified Lagrangian. This correspondence is complete at the level of the field equations and, as such, it is not limited to specific solutions (or symmetries) but, rather, it is valid for all of them. In a subsequent article \cite{Afonso:2018mxn} we make made explicit
this idea using the case of electromagnetic fields. The main aim of the present work is to implement in detail this procedure for scalar matter fields, discussing also some corrections to the results derived in \cite{Afonso:2018bpv}. The consideration of scalar fields is motivated due the their interest for boson \cite{Macedo:2013jja} and Proca stars \cite{Brito:2015pxa}, rotating black holes \cite{Herdeiro:2018daq}, black hole shadows \cite{Cunha:2018acu}, hairy solutions \cite{Herdeiro:2015waa}, inflation \cite{ArmendarizPicon:1999rj}, accelerating solutions \cite{ArmendarizPicon:2000ah}, or topological defects \cite{Bazeia:2007df}, among many others.

The map between theories that we present here proposes the reinterpretation of the terms on the right-hand-side of the Einstein frame metric field equations in such a way that they take on the same structure as the stress-energy tensor of a nonlinear matter field.  This identification is subject to certain integrability conditions, related to stress-energy conservation, between the effective Lagrangian and its partial derivatives, which involve both the metric field equations and the scalar field ones. By carefully analyzing these conditions we show that the correspondence is always well defined for generic (minimally-coupled) matter Lagrangians made out of the scalar field and its quadratic kinetic term. This means that given an RBG coupled to a scalar field Lagrangian, one can always find a new scalar field Lagrangian coupled to GR whose solutions are in correspondence with those of the original RBG theory. Moreover, the inverse problem is also true, namely, given GR coupled to a  scalar field Lagrangian, it is always possible to obtain the modified scalar Lagrangian coupled to a chosen RBG whose solutions can be generated using those of the GR case. As particular cases of interest, we will see that the map that relates a nonlinear gravity theory coupled to a free canonical scalar field to GR, generically leads to a nonlinear matter Lagrangian\footnote{The nonlinear transformation of the matter Lagrangian was overlooked in \cite{Afonso:2018bpv}. Nonetheless, within the range of parameters considered there, the analytical solutions obtained were in excellent agreement with the numerical results, though they were just approximations rather than exact solutions. }. Conversely, if one starts with GR coupled to a canonical, free scalar field, the map to nonlinear gravity theories also involves a nonlinear realization of the matter sources. We explicitly reconstruct those matter sources in the examples mentioned above. The results of this analysis are particularly useful within the applications of non-canonical scalar fields in the literature, see e.g. \cite{ArmendarizPicon:1999rj,ArmendarizPicon:2000ah,Bazeia:2007df}

The content of this work is organized as follows: in Sec.\ref{sec:II} we establish the general framework for RBGs and derive the corresponding field equations. In Sec.\ref{sec:III} the main elements of the mapping are provided, together with explicit applications to quadratic (Starobinski) $f(R)$ models \cite{Starobinski80} and to the Eddington-inspired Born-Infeld theory of gravity  \cite{BanFer10}, besides a particular example. The extension to $N$-components real scalar fields is carried out in Sec.\ref{sec:ssf}, and illustrated with the same two gravitational theories above. We conclude in Sec.\ref{sec:IV} with a discussion of our results and some perspectives for future research.

\section{Field equations for Ricci-based gravities} \label{sec:II}

In this work we refer to Ricci-based gravities (RBGs) as the family of metric-affine  theories defined by an action of the form
\be\label{eq:f-action}
\mathcal{S}=\int d^4x \sqrt{-g} \LL_G\left(g_{\m\n},{R}_{(\m\n)}(\G)\right)+\mathcal{S}_m[g_{\m\n},\psi_m] \ ,
\ee
where the gravitational Lagrangian $\LL_G\left(g_{\m\n},{R}_{(\m\n)}(\G)\right)$ is a scalar function built out of the spacetime metric $g_{\mu\nu}$ (with $g$ denoting its determinant) and the (symmetrized) Ricci tensor ${R}_{(\m\n)}(\G)$ of the  affine connection $\G^\alpha_{\mu\nu}$, which is {\it a priori} independent of the metric, {\it i.e.} $R_{\mu\nu}\equiv{R^\alpha}_{\mu\alpha\nu}$, where the Riemann tensor is defined as $ {R^\a}_{ \b\m\n}=\p_\m\G_{\n\b}^\a-\p_\n\G_{\m\b}^\a+\G_{\m\l}^\a\G_{\n\b}^\l-\G_{\n\l}^\a\G_{\m\b}^\l$. The matter sector, $\mathcal{S}_m=\int d^4x \sqrt{-g}\mathcal{L}_m(g_{\mu\nu},\psi_m)$, contains the matter fields $\psi_m$ and is only coupled to the metric $g_{\mu\nu}$. A careful and rather complete discussion of the role of torsion (the antisymmetric part of the connection) within the field equations of RBGs was carried out by some of us in \cite{Afonso:2017bxr}. It was shown there that for bosonic fields, which is the case that concerns us in this work, torsion can be set to zero by a gauge choice  related to the projective invariance of these theories (hence the need to symmetrize the Ricci). On practical grounds, therefore, one can just forget about torsional terms and set them to zero at the end of the variation with respect to the connection.

The action (\ref{eq:f-action}) is general enough to encompass a large variety of models previously considered in the literature, including (besides GR itself), $f(R)$, $f(R,R_{\mu\nu}R^{\mu\nu})$, Born-Infeld-inspired theories of gravity, etc. These models correspond, indeed, to structures where  $\LL_G\left(g_{\m\n}, R_{\m\n}\right)$ is an arbitrary scalar function of the object ${M^\m}_\n~\equiv~g^{\m\a} R_{\a\n}$. For the sake of this paper, models involving non-minimally coupled matter fields (such as $f(R,T)$) are out of the analysis. However, some recent results \cite{Afonso:2017bxr,Barrientos:2018cnx} have shown that the formalism can be naturally enlarged to accommodate them.

To derive the field equations for the action (\ref{eq:f-action}) we take independent variations with respect to metric and connection, which yields the two systems of equations \cite{Afonso:2017bxr}
\ben
\frac{\p \LL_G}{\p g^{\m\n}}-\frac12 \LL_G g_{\m\n} = T_{\m\n}  \label{eq:gmn}\\
 \nabla^\G_\m \left(\sqrt{-q} q^{\a\b}\right)=0  \ , \label{eq:dGamma}
\een
where $T_{\m\n}=\frac{-2}{\sqrt{-g}}\frac{\d (\sqrt{-g}\LL_m)}{\d g^{\m\n}}$ is the stress-energy tensor of the matter, and
we have introduced the {\it auxiliary} metric $q_{\m\n}$ defined by
\be\label{eq:defq}
\sqrt{-q}\, q^{\m\n}\equiv 2\k^2 \sqrt{-g}  \frac{\p \LL_G}{\p R_{\m\n}}\,,
\ee
with $q$ its determinant, while $\kappa^2$ is a constant with suitable dimensions (in GR, $\kappa^2=8\pi G$). Note that, by construction, $q_{\m\n}$ inherits the index symmetry of the Ricci tensor. Note also that Eq.(\ref{eq:dGamma}) is fully equivalent to the compatibility condition $\nabla^{\G}_{\a}\,q_{\m\n}=0$, which means that $\Gamma_{\mu\nu}^{\lambda}$ is Levi-Civita with respect to $q_{\mu\nu}$; in other words, the components of $\Gamma_{\mu\nu}^{\lambda}$ are given by the Christoffel symbols of $q_{\mu\nu}$:
\begin{equation}
\Gamma_{\mu\nu}^{\lambda}=\frac{1}{2} q^{\lambda \alpha}(\partial_{\mu}q_{\nu\alpha}+\partial_{\nu}q_{\mu\alpha}-\partial_{\alpha}g_{\mu\nu})
\end{equation}

Now, in RBGs it is always possible to introduce a ``deformation matrix" $\W^{\a}_{\ \b}$, implementing the relation between the original (RBG frame) metric $g_{\m\n}$
and the auxiliary (Einstein frame) metric $q_{\m\n}$, through the algebraic relation
\be\label{eq:qWg}
q_{\mu\nu} =  g_{\mu\alpha} {\Omega^\alpha}_{\nu} \ .
\ee
This matrix depends on the matter fields and (possibly) on the spacetime metric $g_{\mu\nu}$ as well, after working out the relation (\ref{eq:defq}) for the particular RBG chosen. Now, tracing with $q^{\mu\alpha}$ over the metric field equations (\ref{eq:gmn}), using the relation (\ref{eq:qWg}), and suitably rearranging terms, one arrives to the set of Einstein-like equations
\be\label{eq:EEmatrix}
{G^\mu}_\nu (q)= \frac{\k^2}{|\hat{\Omega}|^{1/2}} \left[{T^\mu}_\nu-\left(\LL_G +\tfrac{T}{2}\right){\delta^\mu}_\nu\right]\ .
\ee
where ${G^\mu}_{\nu}(q) \equiv q^{\mu\alpha}G_{\alpha\nu}(q)=q^{\mu\alpha} (R_{\mu\nu}(q)-\frac{1}{2}q_{\mu\nu}R(q))$ is the Einstein tensor of the auxiliary metric $q_{\mu\nu}$, $\vert \hat{\Omega} \vert$ denotes the determinant of the matrix ${\Omega^\mu}_{\nu}$, and $T \equiv g^{\mu\nu}T_{\mu\nu}$ is the trace of the stress-energy tensor. Remarkably, the right-hand-side of \eqref{eq:EEmatrix} is completely determined by the matter sources (as $\LL_G$ and $|\hat{\Omega}|$ are on-shell functions of ${T^\mu}_\nu$) and the metric $g_{\mu\nu}$ (generically contained in ${T^\mu}_\nu$). Thus, Eqs.\eqref{eq:EEmatrix} allow to mimic the GR philosophy of having the geometric part on the left-hand-side and the matter contribution on the right-hand-side. Written this way, the effect of the modified dynamics of the RBGs is to engender nonlinearities in the matter sector. Let us point out that in vacuum (${T^\mu}_{\nu}=0$), Eqs.\eqref{eq:EEmatrix} boil down to Einstein equations (with possibly a cosmological constant term), which implies that there are no new dynamical degrees of freedom in these theories. Therefore, in vacuum RBGs only propagate the two tensorial perturbations of the gravitational field travelling at the speed of light, thus allowing these theories to naturally pass the constrains following the almost simultaneous observation of the GW170817 and GRB170817 events \cite{AbbottNS}.

\section{Mapping RBGs with scalar matter into GR} \label{sec:III}

To fix ideas, let us consider a free matter real scalar field, $\mathcal{L}_m=-\frac{1}{2} g^{\alpha\beta}\partial_{\alpha}\phi\partial_{\beta}\phi$, whose stress-energy tensor takes the form
\begin{equation}
{T^\mu}_\nu=g^{\mu\alpha}\partial_\alpha\phi \partial_\nu \phi-\tfrac{1}{2}{\delta^\mu}_\nu g^{\alpha\beta}\partial_\alpha\phi \partial_\beta \phi \ .
\end{equation}
The right-hand-side of Eq.(\ref{eq:EEmatrix}) then becomes
\begin{equation}\label{eq:GmnScalarNaive}
{G^\mu}_\nu (q)= \frac{\k^2}{|\hat{\Omega}|^{1/2}} \left[g^{\mu\alpha}\partial_\alpha\phi \partial_\nu \phi-\LL_G{\delta^\mu}_\nu\right] \ .
\end{equation}
In the case of $f(R)$ theories, for instance, one has $\LL_G=f(R)/2\kappa^2$, thus Eqs.(\ref{eq:defq}) and (\ref{eq:qWg}) yield ${\Omega^\mu}_\nu=f_R {\delta^\mu}_\nu$, while the trace of (\ref{eq:gmn}) provides $R=R(T)$, via the algebraic equation $R f_R-2f=\kappa^2T$, where $f_R \equiv df/dR$ and $T=-g^{\alpha\beta}\partial_\alpha\phi \partial_\beta \phi$. Since by Eq.(\ref{eq:qWg}) one finds $q_{\mu\nu}=f_R g_{\mu\nu}$ and $f_R$ is a function of $T$, which depends on $g_{\alpha\beta}$, it is nontrivial to express $g_{\mu\nu}$ as a function of the matter sources and $q_{\mu\nu}$, though it is possible. In fact, by noting that $\tilde{T}\equiv-q^{\alpha\beta}\partial_\alpha\phi \partial_\beta\phi=T/f_R(T)$, one could invert this relation to obtain $T=T(\tilde{T})$ and write the right-hand-side of (\ref{eq:GmnScalarNaive}) in terms of $q_{\mu\nu}$ and the first derivatives of $\phi$ contracted with $q^{\alpha\beta}$. The corresponding result would take the form
\begin{equation}
\label{eq:GmnScalarNaive2}
{G^\mu}_\nu (q)= \kappa^2 \left[\frac{1}{f_R(R[\tilde{T}])}q^{\mu\alpha}\partial_\alpha\phi \partial_\nu \phi-\frac{f(R[\tilde{T}])}{2\kappa^2f_R^2}{\delta^\mu}_\nu\right] \ .
\end{equation}
This expression suggests that the right-hand-side could be written as the stress-energy tensor of a scalar field with a nonlinear Lagrangian of the form $K(Z)$ with $Z=q^{\alpha\beta}\partial_\alpha\phi \partial_\beta\phi$.
That idea is further reinforced by the fact that the contracted Bianchi's identities $\nabla^{(q)}_{\mu} {G^{\mu}}_{\nu}=0$, impose the conservation
of the right-hand-side as well, which is automatically accomplished if it takes the form
\begin{equation} \label{eq:Kscalar}
{\tilde{T}^\mu}_{\ \ \nu}=K_Z q^{\mu\alpha}\partial_\alpha\phi\partial_\nu\phi-\frac{K(Z,\phi)}{2}{\delta^\mu}_{\nu} \ .
\end{equation}
As we will see next, this can be rigorously formalized for arbitrary RBGs coupled to generic scalar matter Lagrangians.

\subsection{General form of the mapping} \label{sec:IIIA}

When the gravity Lagrangian is more general than the $f(R)$ case, the relation between the spacetime, $g_{\mu\nu}$, and auxiliary, $q_{\mu\nu}$, metrics takes the form of Eq.(\ref{eq:qWg}), with ${\Omega^\mu}_\nu$ being a nonlinear function of the ${T^\mu}_\nu$ of the matter sources. For a real scalar field with a generic non-canonical action
\begin{equation} \label{eq:scalarRBG}
\mathcal{S}_m(X,\phi)=-\frac{1}{2}\int d^4x\sqrt{-g}P(X,\phi) \ ,
\end{equation}
where $X=g^{\alpha\beta}\partial_\alpha\phi\partial_\beta\phi$ and $P$ is some arbitrary function of its arguments, the stress-energy tensor reads
\begin{equation}\label{eq:TmnX}
{{T}^\mu}_\nu=P_X g^{\mu\alpha}\partial_\alpha\phi\partial_\nu\phi- \frac{P(X,\phi)}{2}{\delta^\mu}_\nu \ ,
\end{equation}
where $P_X \equiv dP/dX$. In this case, one can formally consider a series expansion for ${\Omega^\mu}_\nu$ of the form
\begin{equation}
{\Omega^\mu}_\nu=a_0(X,\phi){\delta^\mu}_{\nu}+a_1(X,\phi){T^\mu}_\nu+a_2(X,\phi){T^\mu}_\alpha{T^\alpha}_\nu+\ldots
\end{equation}
A crucial simplifying property of the above expansion arises when one writes (\ref{eq:TmnX}) as ${T^\mu}_\nu=P_X {X^\mu}_\nu-{\delta^\mu}_\nu P/2$, with ${X^\mu}_\nu\equiv g^{\mu\alpha}\partial_\alpha\phi\partial_\nu\phi$ (so that $X$ is simply its trace) and one notes that all powers of $\hat X\equiv {X^\mu}_\nu$  turn out to be proportional to itself, thus leading to $\hat X^n=X^{n-1}\hat X$. As a result, ${\Omega^\mu}_\nu$ must necessarily have the form
\begin{equation}\label{eq:OmegaX}
{\Omega^\mu}_\nu=C(X,\phi){\delta^\mu}_{\nu}+D(X,\phi){X^\mu}_\nu \ ,
\end{equation}
 where $C(X,\phi)$ and $D(X,\phi)$ are model-dependent functions. Thanks to this structure, it is now possible to show that the dependence of the right-hand-side of Eq.(\ref{eq:EEmatrix}) on $g^{\mu\nu}$ can be completely eliminated in favor of $q^{\mu\nu}$ and the matter fields. To see this, note that from the relations (\ref{eq:qWg}), it follows that $g^{\mu\alpha}\partial_\alpha\phi=q^{\mu\alpha}{\Omega_\alpha}^\lambda\partial_\lambda\phi$. Given that Eq.(\ref{eq:OmegaX}) implies ${\Omega_\alpha}^\lambda\partial_\lambda\phi=(C+DX)\partial_\alpha\phi$, from Eq.(\ref{eq:qWg}) we find that
\begin{equation}\label{eq:Xmn2Zmn}
{X^\mu}_\nu=(C+DX){Z^\mu}_\nu \  \Rightarrow \ Z=\frac{X}{C+DX} \ ,
\end{equation}
where ${Z^\mu}_\nu=q^{\mu\alpha}\partial_\alpha\phi \partial_\nu\phi$, consistently with our definition above of $Z\equiv {Z^\mu}_\mu$. We thus see that $Z=Z(X,\phi)$ can, in principle,  be used to obtain an expression for $X=X(Z,\phi)$. This means that the right-hand-side of Eq.(\ref{eq:EEmatrix}) could be written as the stress-energy tensor of the scalar field theory defined by
\begin{equation} \label{eq:scalarGR}
\tilde{\mathcal{S}}_m(Z,\phi)=-\frac{1}{2}\int d^4x\sqrt{-q}K(Z,\phi) \ ,
\end{equation}
which takes the form
\begin{equation}\label{eq:TmnZ}
{\tilde{T}^\mu}_{\ \ \nu}=K_Z {Z^\mu}_\nu- \frac{K(Z,\phi)}{2}{\delta^\mu}_\nu \ .
\end{equation}
Thus, in order to establish the mapping between the RBG coupled to the scalar matter action (\ref{eq:scalarRBG}) and GR coupled to another scalar field (\ref{eq:scalarGR}), one must solve the algebraic equation
\begin{eqnarray}\label{eq:themap}
{{\tilde T}^\m}_{\ \ \nu}& =&\frac{1}{|\hat{\Omega}|^{1/2}} \left[{T^\mu}_\nu-\left(\LL_G +\tfrac{T}{2}\right){\delta^\mu}_\nu\right] \ , \\
&=& \frac{1}{|\hat{\Omega}|^{1/2}} \left[P_X{X^\mu}_\nu-\left(\LL_G +\frac{X P_X-P}{2}\right){\delta^\mu}_\nu\right] \ , \nonumber
\end{eqnarray}
and also verify that the solution is compatible with the evolution equation of the scalar field. That equation comes from variation of the matter action with respect to the scalar field and can be written in the two equivalent forms, namely
\begin{equation}
\partial_\mu\left(\sqrt{-g}P_Xg^{\mu\alpha}\partial_\alpha\phi\right)-\sqrt{-g}\frac{P_\phi}{2}=0 \ ,
\end{equation}
and
\begin{equation}
\partial_\mu\left(\sqrt{-q}K_Zq^{\mu\alpha}\partial_\alpha\phi\right)-\sqrt{-q}\frac{K_\phi}{2}=0 \ .
\end{equation}
Now, attending to the fact that, from Eqs.(\ref{eq:qWg}) and (\ref{eq:Xmn2Zmn}), one has  \begin{equation}
\sqrt{-g}P_Xg^{\mu\alpha}\partial_\alpha\phi=\sqrt{-q}|\hat{\Omega}|^{-1/2}P_X (C+DX)q^{\mu\alpha}\partial_\alpha\phi
\end{equation}
then it follows that, together with Eq.(\ref{eq:Xmn2Zmn}), the relevant relations of this mapping are
\begin{eqnarray}
K(Z,\phi)&=& \frac{1}{|\hat{\Omega}|^{1/2}}\left(2\LL_G +{X P_X-P}\right) \label{eq:diag}\\
K_Z {Z^\mu}_\nu &=& \frac{P_X{X^\mu}_\nu}{|\hat{\Omega}|^{1/2}} \ \ \Rightarrow \ \ ZK_Z=\frac{XP_X}{|\hat{\Omega}|^{1/2}} \label{eq:offdiag} \\
K_\phi&=&\frac{ P_\phi}{|\hat{\Omega}|^{1/2}} \ . \label{eq:potentials}
\end{eqnarray}
Note that Eqs.(\ref{eq:diag}) and (\ref{eq:offdiag}) arise from identifying the diagonal and non-diagonal terms on both sides of Eq.(\ref{eq:themap}), respectively. The second of those equations is consistent with the scalar field equation, which demands
\begin{equation}\label{eq:Kz}
K_Z=\frac{P_X(C+DX)}{|\hat{\Omega}|^{1/2}}  \ .
\end{equation}
If relation (\ref{eq:Xmn2Zmn}) can be inverted to obtain $X=X(Z,\phi)$, then the $K(Z,\phi)$ Lagrangian follows automatically by inserting that expression into Eq.(\ref{eq:diag}). The consistency of this approach requires that the partial derivatives of $K(Z,\phi)$ obtained from Eq.(\ref{eq:diag}) should agree with the expressions given in (\ref{eq:potentials}) and  (\ref{eq:Kz}). In this respect, it is worth noting that, in general, it is much easier to find an expression for $K(Z,\phi)$ in terms of the variables $X$ and $\phi$ than inverting the relation (\ref{eq:Xmn2Zmn}) to write explicitly $K=K(Z,\phi)$. As a result, from a practical point of view it will be much more convenient to compute $K_X$ directly from (\ref{eq:diag}) and compare it with $K_Z Z_X$, where $K_Z$ comes from (\ref{eq:Kz}) and $Z_X$ follows automatically from (\ref{eq:Xmn2Zmn}). For the verification of (\ref{eq:potentials}), one should note that $K_\phi$ actually denotes $\partial_\phi K(Z,\phi)$, which can be written as
\begin{equation}
\partial_\phi K(Z,\phi)=\partial_\phi K(X,\phi)-K_Z Z_\phi \ ,
\end{equation}
with $K_Z$ given in (\ref{eq:Kz}) and $Z_\phi$ computable using (\ref{eq:Xmn2Zmn}).

\subsection{$f(R)$ theories } \label{sec:fR}

As pointed out above, for $f(R)$ theories  we have that $\mathcal{L}_G=f(R)/2\kappa^2$ and ${\Omega^\mu}_\nu=f_R {\delta^\mu}_\nu$ which, from the general expression (\ref{eq:OmegaX}), leads to $C(X,\phi)=f_R$ and $D(X,\phi)=0$. In these theories $f_R$ must be seen as a function of the matter fields alone. Its explicit dependence follows from solving $R f_R-2f=\kappa^2T=\kappa^2(X P_X-2P)$ for a specific function $f(R)$. From Eq.(\ref{eq:diag}) the $K=K(X,\phi)$ Lagrangian in this case is thus given by
\begin{equation}\label{eq:Kf(R)a}
K(X,\phi)=\frac{1}{f_R^2}\left[\frac{f}{\kappa^2}+XP_X-P\right] \ .
\end{equation}
If the relation $Z=X/f_R$ can be inverted to yield $X=X(Z,\phi)$, it is immediate to obtain $K=K(Z,\phi)$ as $K=K(X(Z),\phi)$.

For illustrative purposes, let us consider the quadratic gravity model
\begin{equation} \label{eq:fRquadratic}
f(R)=R+\alpha R^2 \ ,
\end{equation}
where $\alpha$ is a constant with dimensions of length squared. For this model one has $R=-\kappa^2T=-\kappa^2(XP_X-2P)$.  It is immediate to verify by direct calculation that, for any Lagrangian $P(X,\phi)$, the $K(Z,\phi)$ Lagrangian obtained in Eq.(\ref{eq:diag}) is consistent with the conditions imposed by the partial derivatives in Eqs.(\ref{eq:potentials}) and (\ref{eq:Kz}). Consider now for simplicity a (canonical) scalar Lagrangian of the form
\begin{equation}
P(X,\phi)=X-2V(\phi) \ ,
\end{equation}
for which $R=\kappa^2(X-4V(\phi))$. Then, from Eq.(\ref{eq:Xmn2Zmn}) one finds
\begin{equation}
X=\frac{Z(1-8\alpha\kappa^2V(\phi))}{1-2\alpha\kappa^2Z} \ ,
\end{equation}
which inserted in Eq.(\ref{eq:Kf(R)a}) yields the Lagrangian density
\begin{equation}\label{eq:Kf(R)b}
K(Z,\phi)=\frac{Z-\alpha\kappa^2Z^2}{1-8\alpha\kappa^2V(\phi)}-\frac{2V(\phi)}{1-8\alpha\kappa^2V(\phi)} \ .
\end{equation}
It is worth noting that in the case of a free scalar field, $V(\phi)=0$, the Einstein-frame scalar Lagrangian density is simply
\begin{equation} \label{eq:quadraticscalar}
K(Z,\phi)=Z-\alpha\kappa^2Z^2 \ ,
\end{equation}
which, aside a sign, nicely mimics the quadratic structure of (\ref{eq:fRquadratic}). The bottom line of this result is that the nonlinearities on the gravitational sector (since we started with a canonical scalar field Lagrangian density) have been transferred to the matter sector via this correspondence. Later in this section we shall show an explicit example where this procedure is implemented to generate new solutions.

\subsubsection{Inverse problem: obtaining $P(X,\phi)$}

It is also possible to consider the inverse problem, {\it i.e.}, starting from GR coupled to some scalar field described by $K=K(Z,\phi)$, to generate the Lagrangian $P(X,\phi)$, associated to some $f(R)$ theory. To proceed, we take first Eq.(\ref{eq:Kf(R)a}) and use Eq.(\ref{eq:offdiag}) to put it in the form
\begin{equation}\label{eq:Xf(R)}
P(Z,\phi)=\frac{f(R)}{\kappa^2} +f_R^2\left[Z K_Z-K\right] \ .
\end{equation}
The next step requires finding an expression for $R$ 
as a function of $K(Z,\phi)$.
As $g^{\mu\nu}=f_R\, q^{\mu\nu}$, we have $R=g^{\mu\nu}R_{\mu\nu}(q)=f_R\, q^{\mu\nu} R_{\mu\nu}(q)$ and,
 given that in the Einstein frame $R(q)=-\kappa^2\tilde{T}$, we get
\begin{equation} \label{eq:GRtoRBG}
\frac{R}{f_R}=\kappa^2(2K-ZK_Z) \ .
\end{equation}
This allows to algebraically obtain $R$ as a function of the Einstein frame scalar field Lagrangian.
This is all we needed to obtain $P=P(Z,\phi)$ in Eq.(\ref{eq:Xf(R)}).
One can now make use of Eq.(\ref{eq:Xmn2Zmn}),  which in the $f(R)$ case becomes $X=Z f_R$,
 to find an expression for $Z=Z(X)$, to be used in (\ref{eq:Xf(R)}) to eventually find $P=P(X,\phi)$.

To illustrate the procedure above, let us take for simplicity the canonical scalar field Lagrangian (on the Einstein frame)
\begin{equation} \label{eq:canGR}
K(Z,\phi)=Z-2V(\phi) \ .
\end{equation}
Inserting it in Eq.(\ref{eq:GRtoRBG}) one finds
\begin{equation}\label{eq:ZGR}
P(X,\phi)=\frac{X+\alpha\kappa^2X^2}{1+8\alpha\kappa^2V(\phi)}-\frac{2V(\phi)}{1+8\alpha\kappa^2V(\phi)}  \ .
\end{equation}
Comparison of this result with Eq.(\ref{eq:Kf(R)b}) indicates that going from the Einstein frame with a canonical field to the $f(R)$ frame induces a transformation on the scalar Lagrangian which is formally equivalent to that occurring when a canonical field is transformed from the $f(R)$ frame to the Einstein frame, see Eqs.(\ref{eq:fRquadratic}) and (\ref{eq:quadraticscalar}). The only difference is a sign in the parameter that controls the nonlinearity in the gravitational sector. Needless to say that mapping the matter Lagrangian (\ref{eq:ZGR}) back to the Einstein frame one recovers the original matter Lagrangian (\ref{eq:canGR}).

\subsection{Eddington-inspired Born-Infeld gravity theory}

Let us now consider the Eddington-inspired Born-Infeld  (EiBI) gravity theory, whose many applications have been extensively discussed in the literature in the last few years \cite{Banados,BI1,BI2,BI4,BI5,BI5b,BI6,BI7,Gu:2018lub,BI8,Bouhmadi-Lopez:2018sto} (see \cite{BeltranJimenez:2017doy} for a comprehensive review on this kind of theories). In this case, the gravitational action is given by
\begin{equation}  \label{eq:actionEiBI}
\mathcal{S}_{EiBI}=\frac{1}{\kappa^2 \epsilon} \int d^4 x \left[\sqrt{-\vert g_{\mu\nu} + \epsilon R_{\mu\nu}\vert}-\lambda \sqrt{-g}\right] \ ,
\end{equation}
A perturbative expansion in the (length-squared) parameter $\epsilon$ for fields $\vert R_{\mu\nu} \vert \ll 1/\epsilon$ on the action above yields GR+ $\Lambda_{eff}+\mathcal{O}(\epsilon)$, where the effective cosmological constant is given by $\Lambda_{eff}=\frac{\lambda-1}{\epsilon \kappa^2}$. Therefore, in EiBI gravity, deviances with respect to GR solutions occur only in high-curvature (or high-energy density) environments, thus being safe, for instance, from the point of view of the reported equality of the speed of propagation of gravitational waves and electromagnetic radiation in vacuum, see \cite{BI5c} for a discussion on this point.

In this case, the relation between ${\Omega^\mu}_\nu$ and the matter fields is given by
\begin{equation}
|\hat{\Omega}|^{\frac{1}{2}}{[\Omega^{-1}]^{ \mu}}_\nu=\lambda {{\delta^\mu}_{\nu}}-\epsilon\kappa^2{T^\mu}_\nu \ .\label{eq:Om-EiBI}
\end{equation}
For a scalar field with the ${T^\mu}_\nu$ given in Eq.(\ref{eq:TmnX}), this relation takes the form
\begin{equation}\label{eq:Omega-1BIX}
|\hat{\Omega}|^{\frac{1}{2}}{[\Omega^{-1}]^{ \mu}}_\nu= A\, {\delta^\mu}_{\nu}+B{X^\mu}_\nu \ ,
\end{equation}
where we have defined the functions
\begin{eqnarray}
A\equiv A(X,\phi)&=& \lambda+\frac{\epsilon\kappa^2}{2}P(X,\phi) \\
B\equiv B(X,\phi)&=& -\epsilon\kappa^2P_X \ .
\end{eqnarray}
Now, Eq.(\ref{eq:Omega-1BIX}) can be inverted as
\begin{equation}
{\Omega^{\mu}}_\nu =  C{\delta^\mu}_{\nu}+D{X^\mu}_\nu \ ,
\end{equation}
with the definitions
\begin{eqnarray}
C\equiv C(X,\phi)&=& \sqrt{A(A+BX)}\\
D\equiv D(X,\phi)&=& -B\sqrt{\frac{A}{A+BX}} \ ,
\end{eqnarray}
so that the determinant of ${\Omega^\mu}_{\nu}$ reads $|\hat{\Omega}|= A^3(A+XB)$.

It is also useful to have expressions for the above quantities in terms of the $K(Z,\phi)$ Lagrangian of the Einstein frame. These can be obtained using Eq.(\ref{eq:Omega-1BIX}) in combination with (\ref{eq:diag}) and (\ref{eq:offdiag}), together with the fact that the EiBI Lagrangian density in the action (\ref{eq:actionEiBI}) can be written under the compact form\footnote{This is so because in EiBI gravity the connection-compatible metric $q_{\mu\nu}$ turns out to be $q_{\mu\nu}=g_{\mu\nu} + \epsilon R_{\mu\nu }$. Thus, the first term in the EiBI action (\ref{eq:actionEiBI}) is just $\sqrt{-q}$ which, via the basic definition (\ref{eq:qWg}), leads to the above form of the Lagrangian density.} $\LL_{EiBI}=(|\hat{\Omega}|^{\frac{1}{2}}-\lambda)/\epsilon\kappa^2$. One then finds
\begin{eqnarray}\label{eq:Omega-1BIY}
{[\Omega^{-1}]^{ \mu}}_\nu&=& \tilde{A}\, {\delta^\mu}_{\nu}+ \tilde{B} {Z^\mu}_\nu\\
\tilde A&\equiv& \tilde A(Z,\phi)= 1-\frac{\epsilon\kappa^2}{2}(K-Z K_Z) \q \\
\tilde B&\equiv& \tilde B(Z,\phi)= -\epsilon\kappa^2K_Z \ ,
\end{eqnarray}
which is inverted as
\begin{eqnarray}
{\Omega^{\mu}}_\nu &=&  \tilde C{\delta^\mu}_{\nu}+\tilde D{Z^\mu}_\nu  \label{eq:OmZ}  \\
\tilde C&\equiv& \tilde C(Z,\phi)= \frac{1}{\tilde A}\\
\tilde D&\equiv& \tilde D(Z,\phi)= -\frac{\tilde B}{\tilde A(\tilde A+\tilde B Z)} \ ,
\end{eqnarray}
so that the determinant now becomes $|\hat{\Omega}|= \tilde C^3(\tilde C+Z \tilde D)$. By direct calculation one can verify that for generic $P(X,\phi)$ matter models, the relations above together with the mapped Lagrangian (\ref{eq:diag}) and its partial derivatives (\ref{eq:offdiag}) and (\ref{eq:potentials}) are fully consistent.

As an example, let us consider the EiBI theory (\ref{eq:actionEiBI}) coupled to the family of scalar field models defined by
\begin{equation}
P(X,\phi)=p(X)-2V(\phi) \ ,
\end{equation}
The corresponding scalar Lagrangian in the Einstein frame can be written in parametric form as
\begin{eqnarray}
Z(X,\phi)&=& 2 X \sqrt{\frac{a}{b^3}}\\
K(X,\phi)&=&  \frac{2(\sqrt{a b^3}-2(a+\epsilon\kappa^2 Xp_X))}{\epsilon\kappa^2\sqrt{a b^3}}\\
\text{with           }\qq&&\notag\\
a&\equiv & 2\lambda+\epsilon\kappa^2[ p(X)-2 V(\phi)-2  X p_X]  \qq\qq\\
b&=& 2\lambda+\epsilon\kappa^2[ p(X)-2 V(\phi)] \ .
\end{eqnarray}
If one focuses, for simplicity, on the free canonical scalar case, $\{p(X)=X; V=0\}$, the above expressions become explicitly
\begin{eqnarray}
Z(X)&=& X \left(\lambda-\tfrac{\epsilon\kappa^2 }{2}X\right)^{\!\frac12}\!\left(\lambda+\tfrac{\epsilon\kappa^2 }{2}X\right)^{-\!\frac32}\\
K(X)&=&\! \frac{2}{\epsilon\kappa^2}\!\left(1-{\lambda}{
\left(\lambda-\tfrac{\epsilon\kappa^2 }{2}X\right)^{-\frac12}\!\!\left(\lambda+\tfrac{\epsilon\kappa^2 }{2}X\!\right)^{-\frac32}}\right). \q
\end{eqnarray}
The weak-field expansion of the above expressions leads to
\begin{equation}
K(Z)\approx Z+\frac{\epsilon\kappa^2}{4}Z^2 \ ,
\end{equation}
whereas the strong-field regime depends on the sign of $\epsilon$. If $\epsilon>0$, in that region we find that beyond the threshold $X_{Max}=1/\epsilon\kappa^2$ the Lagrangian density $K(Z)$ is no longer a real function. The linear approximation turns out to be a good one all over this domain, which ends at $Z_{Max}=(2/\sqrt{27})/\epsilon\kappa^2$. If $\epsilon<0$, the domain of $X$ is bounded, with $X_{Max}=\vert 2/\epsilon\kappa^2 \vert $, but $Z$ is unbounded from above. In that asymptotic limit, one finds that $K(Z)\approx (\vert 2/\epsilon\kappa^2 \vert +Z/2+(3/2)Z^{1/3})/ \vert \epsilon\kappa^2 \vert^{2/3}$ is a very good approximation. The fact that the linear term dominates in this regime justifies the excellent agreement between the analytical approximation found in \cite{Afonso:2018bpv} and the numerical results of \cite{Afonso:2017aci}.

\subsubsection{Inverse problem}

Let us now consider the problem of mapping a scalar field matter model coupled to GR into another scalar field model coupled to the EiBI gravity.
Following similar steps as in the $f(R)$ case above, it is easy to show that, given a scalar theory $K(Z,\phi)$ and a gravitational Lagrangian $\LL_G$,
it is always possible to find the associated scalar field Lagrangian $P=P(Z,\phi)$ using the combination of Eqs.(\ref{eq:diag}) and (\ref{eq:offdiag}) as
\begin{equation}\label{eq:PZBIa}
P(Z,\phi)=2\LL_G+|\hat{\Omega}|^{1/2}(ZK_Z-K) \ .
\end{equation}
According to expressions for $|\hat{\Omega}|$ and $\LL_{G}$ in terms of $Z$ for the EiBI theory given in the previous section, one can easily find
\begin{equation}\label{eq:PZGen}
P(Z,\phi)=\frac{2}{\epsilon \kappa ^2} \left(\tfrac{2}{\sqrt{\left[2+\epsilon \kappa ^2 \left(K+Z K_Z\right)\right] \left[2+\epsilon \kappa ^2 \left(K-Z K_Z\right)\right]}}-\lambda\right)\ .
\end{equation}

In general, for RBGs, one must note that in order to get a scalar object, the nonlinear function $\LL_{G}$ must be made out of traces of powers of the object $g^{\mu\alpha}R_{\alpha\nu}(\Gamma)$.
As, on shell, $R_{\alpha\nu}(\Gamma)=R_{\alpha\nu}(q)$, one can thus use the relation $g^{\mu\alpha}R_{\alpha\nu}(q)={\Omega^\mu}_\alpha q^{\alpha\beta}R_{\beta\nu}(q)$
and the fact that in the Einstein frame $q^{\alpha\beta}R_{\beta\nu}(q)=\kappa^2({\tilde{T}^\alpha}_\nu-\tilde{T}\delta^\alpha_\nu/2)$,
to express $g^{\mu\alpha}R_{\alpha\nu}(\Gamma)$ in terms of quantities related to $K(Z,\phi)$.
On the other hand, since ${\Omega^\mu}_\alpha$ must be of the form given in Eq.(\ref{eq:OmZ}), Eq.(\ref{eq:Xmn2Zmn})
can also be written as $X=(\tilde C+\tilde D Z)Z$, thus providing a parametric representation for $P(X,\phi)$.

As an example of the above reasoning, lets investigate how the canonical matter model
\begin{equation}
K(Z,\phi)=Z-2V(\phi) \ ,
\end{equation}
coupled to GR gets mapped to the EiBI framework. Inserting this Lagrangian in Eq.(\ref{eq:PZBIa}) and using the expression of the determinant for EiBI gravity obtained above, we get
\begin{equation}\label{eq:PZBIb}
P(Z,\phi)=\frac{2}{\epsilon \kappa ^2} \!\left(\tfrac{1}{\sqrt{\left(1+\epsilon \kappa ^2 V(\phi )\right) \left(1-Z \epsilon \kappa ^2+\epsilon \kappa ^2 V(\phi )\right)}}-\lambda\right)\ ,
\end{equation}f
while the relation between $Z$ and $X$ becomes
\begin{equation}\label{eq:XofZBI}
Z=\frac{X (1+\epsilon\kappa^2 V(\phi ))}{1+\epsilon\kappa^2 X} \ .
\end{equation}
Combining these two expressions, one obtains
\begin{equation}\label{eq:PZBIc}
P(X,\phi)=\frac{2 \left(\sqrt{1+ \epsilon \kappa ^2 X}-\lambda(1+\epsilon \kappa ^2 V(\phi))\right)}{\epsilon \kappa ^2 \left(1+\epsilon \kappa ^2 V(\phi )\right)} \ ,
\end{equation}
The free field case, $V(\phi)\to 0$, reduces to
\begin{equation}\label{eq:PZBId}
P(X)=\frac{2 \left(\sqrt{1+ \epsilon \kappa ^2 X}-\lambda\right)}{\epsilon \kappa ^2} \ ,
\end{equation}
which has the characteristic square-root structure of Born-Infeld-like theories of matter \cite{BI34,Felder:2002sv,Jana:2016uvq}. One can verify that this Lagrangian in the EiBI frame transforms into its original form $Z-2V(\phi)$ in the Einstein frame, which confirms the consistency of our approach.

\subsection{Generating exact solutions}

Before moving forward, let us further illustrate the power of the above developed method as a tool for constructing exact analytical solutions for RBGs.
For this purpose, we take a static spherically symmetric free scalar field in GR, and use it to generate the corresponding solution
in the quadratic $f(R)$ model discussed above (see section \ref{sec:fR} and Eq. (\ref{eq:fRquadratic})).
The solutions for this GR scalar field model were originally obtained by Wyman in Ref.\cite{Wyman}.
Since in static spherically symmetric spacetimes there are only two nontrivial independent metric functions, the line element can be suitably cast into the form
\begin{equation}
ds_{GR}^2=-e^{\nu}dt^2+\frac{e^\nu}{W^4}dy^2 +\frac{1}{W^2}(d\theta^2+\sin\theta^2d\varphi^2)  \ ,
\end{equation}
where $\nu$ and $W$ are functions of the radial coordinate $y$. This unusual form of the line element is justified on the fact that it leads to a very simple equation for the scalar field, namely, $\phi_{yy}=0$. Without loss of generality, its solution can be taken as $\phi=y$. Demanding asymptotic flatness, the solutions of Einstein's equations take the form \cite{Wyman}
\begin{eqnarray}
e^\nu &=& e^{\beta y} \\
W &=& \gamma^{-1} e^{\beta y/2}\sinh(\gamma y) \ ,
\end{eqnarray}
where the constant $\beta$ is related to the asymptotic Newtonian mass of the solution as $\beta =-2GM$, and we have introduced the constant $\gamma\equiv \sqrt{\beta^2+2\kappa^2}/2$.

Since this solution corresponds to GR coupled to the canonical Lagrangian $K(Z)=Z$, with it we can generate the corresponding solution in many different RBGs. For instance, in the case of the quadratic $f(R)=R+\alpha R^2$ model (\ref{eq:fRquadratic}), the scalar Lagrangian $K(Z)=Z$ is mapped into $P(X)=X+\alpha \kappa^2 X^2$, as shown in Eq.(\ref{eq:ZGR}). Since the scalar field profile is already known, we just need to write the relation between the metrics in terms of quantities obtained in the Einstein frame. This means, in particular, that the function $f_R$ that relates $g_{\mu\nu}$ and $q_{\mu\nu}$ must be written in terms of $Z$ rather than $X$. Using the relation (\ref{eq:GRtoRBG}), it is easy to see that
\begin{equation}
f_R=\frac{1}{1-2\alpha \kappa^2 Z} \ ,
\end{equation}
where $Z=q^{\mu\nu}\partial_\mu\phi \partial_\nu\phi =q^{yy}\phi_y^2=q^{yy}=W^4 e^{-\nu}$. Therefore, the line element corresponding to $g_{\mu\nu}$ becomes
 \begin{equation}
ds_{f(R)}^2=\frac{1}{1-2\alpha \kappa^2 W^4 e^{-\nu}} ds_{GR}^2 \ ,
\end{equation}
Had we chosen the EiBI theory as our modified gravity model, the corresponding line element  would 
take the form
\begin{equation} \label{dsEiBI}
ds_{EiBI}^2=-e^{\nu}dt^2+\left[\frac{e^\nu}{W^4}-\epsilon \kappa^2\right]dy^2 +\frac{1}{W^2}(d\theta^2+\sin\theta^2d\varphi^2)  \ .
\end{equation}

Despite of the innocent appearance of the above line element \eqref{dsEiBI}, a quick analysis puts forward a dramatic modification of the physical content as compared to GR.
For instance, the radial proper distance in the GR case as one approaches the central region $y\to\infty$, shortens exponentially fast, $l_{GR}\propto e^{-\lambda y}$, with $\lambda= \beta+2\sqrt{2\kappa^2+\beta^2}$. On the contrary, in the EiBI case with $\epsilon<0$,
one can easily verify that, in the same approximation, $l_{EiBI}=\sqrt{|\epsilon|}\kappa y$. 
Thus, in this case the center lays at an infinite proper radial distance and therefore the internal structure of these objects is radically different from the GR Wyman's solution. On the other hand, for EiBI with $\epsilon>0$, there exists a maximum value attainable by the $y$ coordinate, $y_{max}=- \log[{\vert \epsilon \vert \kappa^2}/{(2\kappa^2+\beta^2)^2}]$. The neighbourhood of this region exhibits an interesting geometric structure in which both the $t-y$ and the spherical sectors behave as maximally symmetric subspaces, being the $t-y$ one of de Sitter type while the spherical part has divergent curvature. Radial geodesics can get there in a finite proper time.  

Further details of this derivation and an in-depth analysis of its physical implications will be given elsewhere\footnote{See \cite{Afonso:2017aci} for a direct attack on this problem for the case of Born-Infeld gravity coupled to a canonical free scalar field.}.

\section{Several scalar fields} \label{sec:ssf}

In the previous sections we have considered the case of a single real scalar field. We will now extend those results to the case of an arbitrary number of real scalar fields. A similar approach can also be followed for complex fields. In order to parallel our previous derivation, we will need to adapt the notation to accommodate several scalar fields. Let us thus focus on a generic multi-scalar Lagrangian density defined by
\begin{eqnarray}
P&=&P(X_{ij},\phi_k) \\
{{X_{ij}}^\mu}_\nu&\equiv& g^{\mu\alpha}\partial_\alpha\phi_i\partial_\nu \phi_j  \\
X_{ij}&\equiv& g^{\alpha\beta}\partial_\alpha\phi_i\partial_\beta \phi_j =  X_{ji}\ ,
\end{eqnarray}
where $i,j$ run from $1$ to $N$ scalar fields, and whose stress-energy tensor takes the form
 \begin{equation}\label{eq:TmnXMany}
 {T^\mu}_\nu= \sum_{i,j} P^{ij}{{X_{ji}}^\mu}_\nu-\frac{P}{2}{\delta^\mu}_\nu \ .
 \end{equation}
where $P^{ij} \equiv \partial P/\partial X_{ij}$. Inserting this expression in the RBG field equations (\ref{eq:EEmatrix}), one finds
 \begin{eqnarray}\label{eq:GmnMany}
 {G^\mu}_\nu(q)&=&\frac{\kappa^2}{|\hat{\Omega}|^{\frac{1}{2}}}\left[\sum_{i,j} P^{ij}{{X_{ji}}^\mu}_\nu-\right. \\
 & &\left.\q -\frac{1}{2}\left(2\LL_G+\sum_{m,n} P^{mn}{{X_{nm}}}-P \right){\delta^\mu}_{\nu} \right] \ .\nonumber
 \end{eqnarray}
The right-hand-side of this equation should be equal to
 \begin{equation}
 \kappa^2{\tilde T^\mu}_{\ \ \nu}= \kappa^2\sum_{i,j} K^{ij}{{Z_{ji}}^\mu}_\nu-\frac{\kappa^2}{2}K{\delta^\mu}_{\nu} \ , \label{eq:TmnMany}
 \end{equation}
where ${{Z_{ij}}^\mu}_\nu\equiv q^{\mu\alpha}\partial_\alpha\phi_i\partial_\nu \phi_j $ and $K^{ij} \equiv \partial K/\partial Z_{ij}$. The construction of the Lagrangian $K(Z_{ij},\phi_k)$ is a natural generalization of the approach detailed in the previous section. Indeed, comparing the right-hand-side of (\ref{eq:GmnMany}) with (\ref{eq:TmnMany}), we find
\begin{eqnarray}
&&K(Z_{ij},\phi_k)=|\hat{\Omega}|^{-\frac{1}{2}}\left(2\LL_G+\sum_{m,n} P^{mn}{{X_{nm}}}-P \right) \ , \; \label{eq:K-Many}\\
&&\sum_{i,j} K^{ij}{{Z_{ji}}^\mu}_\nu = |\hat{\Omega}|^{-\frac{1}{2}}\sum_{i,j} P^{ij}{{X_{ji}}^\mu}_\nu\,.\label{eq:Kij-Many}
\end{eqnarray}

The first equation provides a parametric representation of the scalar matter Lagrangian density in terms of quantities in the RBG frame,
once a specific solution of the metric field equations for the RBG theory is specified.
To implement this, we expand the $\Omega$-matrix as a power series of the energy-momentum tensor.
Now, the energy-momentum tensor \eqref{eq:TmnXMany} depends  just on two tensorial quantities, namely the identity ${{\d}^\mu}_\nu$ and $X_{ij}{}^\mu{}_\nu$.
These objects reproduce themselves upon multiplications due to the fundamental property
\begin{equation}
X_{ij}{}^\mu{}_\rho X_{mn}{}^\rho{}_\nu = X_{jm}X_{in}{}^\mu{}_\nu\ ,\label{eq:FundProp}
\end{equation}
where $X_{ij}\equiv X_{ij}{}^\mu{}_\mu$.
Therefore, one can propose the general ansatz
\begin{equation}\label{eq:Om-Many}
{[\Omega^{-1}]^\mu}_\nu=A{\delta^\mu}_{\nu}+\sum_{i,j} B^{ij} {{X_{ji}}^\mu}_\nu \ ,
\end{equation}
where $A$ and $B^{ij}$ are model-dependent functions of ${X_{mn}}$ and $\phi_k$,
that can be deduced for each case from the corresponding metric field equations.
The parametrization of the kinetic term $Z_{ij}$ of the scalars in the GR-frame can be established by noticing that
\begin{equation}
Z_{ij}{}^\mu{}_\nu={[\Omega^{-1}]^\mu}_\rho X_{ij}{}^\rho{}_\nu\,,
\end{equation}
where we have used $q^{\mu\alpha}={[\Omega^{-1}]^\mu}_\beta\, g^{\beta\alpha}$. Thus, substituting \eqref{eq:Om-Many} and using \eqref{eq:FundProp}, one finds the generalization of \eqref{eq:Xmn2Zmn}:
\begin{equation}\label{eq:Zij}
Z_{mn}{}^\mu{}_{\rho} = \sum_{i}\left(A\delta^i_m + \sum_{j} B^{ji}X_{mj}\right)X_{in}{}^\mu{}_{\rho}\,.
\end{equation}

The last ingredient needed to find the parametrization of the Lagrangian $K[Z_{ij}(X_{mn},\phi_l),\phi_k]$ from Eq.(\ref{eq:K-Many}) is the determinant of the $\Omega$-matrix.
This problem is reduced to the calculation of traces of powers of it. Using the property \eqref{eq:FundProp}  one can easily find that
\begin{eqnarray}
|\hat\Omega|^{-1} = A^4+A^3 B_1+B_1^4+\frac12 A^2\left(B_1^2-B_2\right)-\frac14B_1^2 B_2 +\nonumber
\\+ 3B_2^2 +\frac13B_1 B_3 + \frac16 A\left(B_1^3-3B_1 B_2 + 2 B_3\right) - \frac14 B_4 \ ,\nonumber
\\
\end{eqnarray}
where
\begin{eqnarray}
&&B_1 = \sum_{ij}B^{ij}X_{ij}\,\qquad  B_2 = \sum_{ijkl}B^{ij}B^{kl}X_{il}X_{jk}\ ,\nonumber\\
&&B_3 = \sum_{ijklmn}B^{ij}B^{kl}B^{mn}X_{in}X_{jk}X_{lm}\ ,\nonumber\\
&&B_4 = \sum_{ijklmnpq}B^{ij}B^{kl}B^{mn}B^{pq}X_{iq}X_{jk}X_{lm}X_{np}\ .
\end{eqnarray}
\subsubsection{Inverse multiscalar problem}
The inverse problem can also be worked out by noting that the equations \eqref{eq:Kij-Many} and (\ref{eq:K-Many}) imply the following expression for the scalar sector of the RBG:
\begin{equation}\label{eq:P-Many}
P(X_{ij},\phi_k)=2\LL_G+|\hat\Omega|^{\frac{1}{2}}\left[\sum_{m,n} K^{mn}{{Z_{nm}}}-K \right] \ .
\end{equation}
Therefore, the parametric representation of the RBG-frame Lagrangian $P\left[X_{ij}\left(Z_{mn},\phi_l\right),\phi_k\right]$ is reduced to finding an expression of the $\Omega$-matrix in terms of quantities of the GR-frame,
in order to write the RBG metric in terms of the GR metric via the fundamental relation $g_{\mu\nu}=q_{\mu\rho}(\Omega^{-1})^\rho{}_\nu$ of Eq.(\ref{eq:qWg}).
Following the same arguments as in the single field case, and taking into account that relation \eqref{eq:FundProp} is now replaced by
\begin{equation}
Z_{ij}{}^\mu{}_\rho Z_{mn}{}^\rho{}_\nu = Z_{jm}Z_{in}{}^\mu{}_\nu\ ,\label{eq:FundPropZ}
\end{equation}
one can write the analogue of \eqref{eq:Om-Many} in the GR-frame, namely
\begin{equation}
{[\Omega^{-1}]^{ \mu}}_\nu = \tilde A{{\delta^\mu}_{\nu}} + \sum_{ij}\tilde B^{ij}Z_{ji}{}^\mu{}_\nu \ .\label{eq:OmOm-Many-Einstein}
\end{equation}
The crucial step of this approach relies on finding the explicit expressions of $\tilde A$ and $\tilde B^{ij}$ in terms of ${Z_{mn}}$ and $\phi_k$,
that depend on the form of the metric field equations and the mapping defining equations \eqref{eq:K-Many} and \eqref{eq:Kij-Many}.
Once $\tilde A$ and $\tilde B^{ij}$ are known, it becomes straightforward to compute the determinant of the $\Omega$-matrix in the GR frame, which reads
\begin{eqnarray}
|\hat\Omega|^{-1} = \left[\tilde A^4+\tilde A^3 \tilde B_1+\tilde B_1^4+\frac12 \tilde A^2\left(\tilde B_1^2-\tilde B_2\right)-\frac14 \tilde B_1^2 \tilde B_2 +\right.\nonumber
\\\left.+ 3\tilde B_2^2 +\frac13\tilde B_1 \tilde B_3 + \frac16 \tilde A\left(\tilde B_1^3-3\tilde B_1 \tilde B_2 + 2 \tilde B_3\right) - \frac14 \tilde B_4\right] \ ,\nonumber
\\
\end{eqnarray}
where
\begin{eqnarray}
&&\tilde B_1 = \sum_{ij}\tilde B^{ij}Z_{ij}\,\qquad  \tilde B_2 = \sum_{ijkl}\tilde B^{ij}\tilde B^{kl}Z_{il}Z_{jk}\ ,\nonumber\\
&&\tilde B_3 = \sum_{ijklmn}\tilde B^{ij}\tilde B^{kl}\tilde B^{mn}Z_{in}Z_{jk}Z_{lm}\ ,\nonumber\\
&&\tilde B_4 = \sum_{ijklmnpq}\tilde B^{ij}\tilde B^{kl}\tilde B^{mn}\tilde B^{pq}Z_{iq}Z_{jk}Z_{lm}Z_{np}\ .
\end{eqnarray}
Furthermore, it is possible to invert explicitly the $\Omega$-matrix representation \eqref{eq:OmOm-Many-Einstein}. Once again, using \eqref{eq:FundPropZ},
one can justify the ansatz
\begin{equation}
\Omega^{\mu}{}_\nu = \tilde C{{\delta^\mu}_{\nu}} + \sum_{ij}\tilde D^{ij}Z_{ji}{}^\mu{}_\nu \ .\label{eq:Om-Einstein}
\end{equation}
The explicit form of $\tilde C$ and $\tilde D^{ij}$ can be obtained by imposing that this expression is the inverse matrix of \eqref{eq:OmOm-Many-Einstein}, leading to the following equations for $\tilde C$ and $\tilde D^{ij}$:
\begin{eqnarray}
&&\tilde A \tilde C = 1\ ,\nonumber\\
&&\tilde C\tilde B^{ij} + \sum_l \tilde D^{il}\left(\tilde A \delta^j_l+\sum_k \tilde B^{kj}Z_{lk}\right) = 0\ .
\end{eqnarray}
The solution can be conveniently expressed introducing matrix notation for latin indices spanning the scalar space
\begin{eqnarray}
\tilde C = \frac{1}{\tilde A}\ ,\quad \hat{\tilde D}^{ij} = -\frac{1}{\tilde A} \left(\tilde A\,I+\hat Z \hat{\tilde B}\right)^{-1}\hat{\tilde B}\ .
\end{eqnarray}
From relation \eqref{eq:Om-Einstein} and the property \eqref{eq:FundPropZ}, one can obtain the parametrization of the kinetic term $X_{mn}(Z_{ij},\phi_l)$ via the following equation
\begin{equation}\label{eq:Xij-Many}
X_{mn}{}^\mu{}_{\rho} = \sum_{i}\left(\tilde C\delta^i_m + \sum_{j} \tilde D^{ji}Z_{mj}\right)Z_{in}{}^\mu{}_{\rho}\,.
\end{equation}


The conditions that the partial derivatives of the Lagrangians $P(X_{ij},\phi_k)$  and $K(Z_{ij},\phi_k) $ must satisfy follow straightforwardly from the equations of motion of the $\phi_k$ scalar fields in the two (RBG and GR, respectively) frames
\begin{eqnarray}
&&\partial_\mu\left[\sqrt{-g} g^{\mu\nu}\left(2P^{ii}\partial_\nu\phi_i+ \sum_{j\neq i} P^{ij}\partial_\nu\phi_j\right)\right] \\
&&-\sqrt{-g}P_{\phi_i}=0 \nonumber \ , \\
&&\partial_\mu\!\left[\sqrt{-q} q^{\mu\nu}\!\left(2K^{ii}\partial_\nu\phi_i + \sum_{j\neq i} K^{ij}\partial_\nu\phi_j\right)\right] \\
&&-\sqrt{-q}K_{\phi_i}=0 \nonumber   \ .
\end{eqnarray}
The most natural compatibility conditions of these two field equations are obtained by matching the spacetime derivative sectors and the derivatives with respect to the scalar fields according to the two equations
\begin{eqnarray}
&&2P^{ii}\partial_\mu\phi_i \!+\! \sum_{j\neq i} P^{ij}\partial_\mu\phi_j \\
&&=|\hat\Omega|^{1/2} \left(\Omega^{-1}\right)^\alpha{}_\mu \times\!\left(2K^{ii}\partial_\alpha\phi_i+ \sum_{j\neq i} K^{ij}\partial_\alpha\phi_j\right) \nonumber \ ,\\ 
&&P_{\phi_i}=|\hat\Omega|^{1/2}K_{\phi_i} \ .
\end{eqnarray}
Using \eqref{eq:Om-Many}, the first
equation provides the following two relations
\begin{eqnarray} \label{eq:Pii}
|\hat\Omega|^{-1/2}P^{ii} &=&\left( A +\sum_k  B^{ki}X_{ki}\right)K^{ii} \notag\\
&+&\frac12 \sum_{k,j\neq i} B^{ki}X_{kj} K^{ij}\,,\\
|\hat\Omega|^{-1/2}P^{ij}&=&\sum_{k\neq i} \left( A \delta^j_k + \sum_m  B^{mj}X_{mk}\right)K^{ik} + \nonumber\\ &+&2 \sum_m  B^{mj}X_{mi} K^{ii}\,. \label{eq:Pij}
\end{eqnarray}

Let us consider the simplest case of two real scalar fields. Then, Eqs.(\ref{eq:Pii})-(\ref{eq:Pij}) reduce to
\begin{eqnarray}
|\hat\Omega|^{-\frac12} P^{11} &=& \left(A + \sum_{m=1}^2 B^{m1}X_{m1}\right)K^{11} \nonumber\\
&+&\frac12 \sum_{m=1}^2 B^{m1}X_{m2} K^{12}
\\
|\hat\Omega|^{-\frac12} P^{22} &=&\left( A +  \sum_{m=1}^2 B^{m2}X_{m2}\right)K^{22} \nonumber\\
&+&\frac12  \sum_{m=1}^2 B^{m2}X_{m1} K^{12}\\
|\hat\Omega|^{-\frac12} P^{12} &=& \left( A +  \sum_{m=1}^2 B^{m2}X_{m2}\right)K^{12} \nonumber\\
 &+& 2 \sum_{m=1}^2 B^{m2}X_{m1} K^{11} \,
\end{eqnarray}
supplemented by the compatibility constraint
\begin{eqnarray}
&&\sum_{m=1}^2\left[\left(B^{m2}X_{m2}- B^{m1}X_{m1}\right)K^{12} + \right.\\
&&\quad\qquad \left.+ 2 \left( B^{m2}X_{m1} K^{11} -  B^{m1}X_{m2} K^{22}\right) \right ]= 0\,. \q\nonumber
\end{eqnarray}

The above expressions generalize the results obtained in Sec.\ref{sec:IIIA} to the case of a set of $N$-components real scalar fields. Next, we shall give two explicit examples illustrating these results.

\subsection{Application 1: $f(R)$ gravity with a complex scalar or many scalar fields}

For its relevance in different astrophysical and cosmological applications, we will next consider a situation involving two real scalar fields coupled to the quadratic $f(R)$ theory discussed in the previous sections.
 Having in mind the case of a complex scalar field, that can be represented in terms of two real fields without the crossed term $X_{12}$, we focus on an action of the form
\begin{equation}
\mathcal{S}_m=-\frac{1}{2}\int d^4x \sqrt{-g}P(X_{11},X_{22};\phi_1,\phi_2) \ .
\end{equation}
We shall furthermore assume the dynamics to be given by two canonical scalar fields, namely
\begin{equation}
P(X_{11},X_{22};\phi_1,\phi_2)=X_{11}+X_{22}-2V(\phi_1,\phi_2) \ .\label{eq:canon-2scal}
\end{equation}
The GR-frame $K(Z_{ij},\phi_k)$ Lagrangian follows directly from Eq.(\ref{eq:K-Many}) and takes the form
 \begin{equation}\label{eq:KofX2fields}
 K=\frac{X_{11}+X_{22}-2V+\alpha \kappa ^2 \left(X_{11}+X_{22}-4V\right){}^2}{\left(1+2 \alpha \kappa ^2 \left(X_{11}+X_{22}-4V\right)\right){}^2} \ .
\end{equation}
The relation between $X_{11}$ and $X_{22}$ with the $Z_{ij}$ is established by Eq.(\ref{eq:Zij}) and leads to
\begin{equation}\label{eq:2scalars}
Z_{11}= \frac{1}{f_R} X_{11} \ \ , \ \ Z_{22}= \frac{1}{f_R} X_{22}  \ ,
\end{equation}
where $f_R=1+2 \alpha \kappa ^2 \left(X_{11}+X_{22}-4V\right)$. Inverting those relations, one finds
\begin{eqnarray}\label{eq:2scalarsA}
X_{11}&=&\frac{Z_{11} \left(1-8 \alpha \kappa ^2 V\right)}{1-2 \alpha \kappa ^2 (Z_{11}+ Z_{22})}\\
X_{22}&=&\frac{Z_{22}}{Z_{11}}X_{11} \ , \label{eq:2scalarsB}
\end{eqnarray}
and inserting this result in Eq.(\ref{eq:KofX2fields}) we finally get
\begin{equation}\label{eq:KofZ2fields}
 K=\frac{\left(Z_{11}+Z_{22}\right) \left(1-\alpha  \kappa ^2 \left(Z_{11}+Z_{22}\right)\right)-2 V}{1-8 \alpha  \kappa ^2 V} \ .
\end{equation}
For free fields, the Lagrangian takes the simpler form
\begin{equation}
 K={\left(Z_{11}+Z_{22}\right) \left(1-\alpha  \kappa ^2 \left(Z_{11}+Z_{22}\right)\right)}\ .
\end{equation}
Denoting $X\equiv X_{11}+X_{22}=g^{\mu\nu}\partial_\mu\phi\,\partial_\nu \phi^*$ and  $Z\equiv Z_{11}+Z_{22}=q^{\mu\nu}\partial_\mu\phi\,\partial_\nu \phi^*$, with $\phi=\phi_1+ i\phi_2$ and $\phi^*=\phi_1-i\phi_2$, it is easy to see that the original RBG canonical scalar Lagrangian $P(X)=X$ turns into the GR non-canonical field $K(Z)=Z-\alpha  \kappa ^2 Z^2$, much in the same way as in the single real scalar field case discussed in Sec.\ref{sec:fR}.

It is straighforward to generalize the above analysis for two real scalar fields to an arbitrary number of scalar fields. In this case the Lagrangian density has the form
\begin{equation}
P(X_{mn},\phi_k)=X_{Tot}-2V(\phi_1,\ldots,\phi_N) \ ,
\end{equation}
where $X_{Tot}=\sum_{m,n} X_{mn} $. The key nontrivial step is the extension of Eq.(\ref{eq:2scalars}), which generically turns into
\begin{equation}\label{eq:Nscalars}
Z_{ij}= \frac{1}{f_R} X_{ij} \ \ , \ \ X_{mn}= \frac{X_{ij}}{Z_{ij}}  Z_{mn} \ ,
\end{equation}
where for $N$ scalar fields we have
\begin{equation}
f_R=1-8\alpha \kappa^2V+2\alpha \kappa^2 X_{Tot} \ .
\end{equation}
Writing in this last expression $X_{Tot}=Z_{Tot} X_{ij}/Z_{ij}$, from the first relation in Eq.(\ref{eq:Nscalars}) one finds
\begin{equation}\label{eq:Xij_N}
X_{ij}=\frac{Z_{ij}(1-8\alpha \kappa^2V)}{1-2\alpha\kappa^2 Z_{Tot}} \ ,
\end{equation}
which also leads to
\begin{equation}\label{eq:XTot_N}
X_{Tot}=\frac{Z_{Tot}(1-8\alpha \kappa^2V)}{1-2\alpha\kappa^2 Z_{Tot}} \ .
\end{equation}
Since the parametric representation of the Lagrangian density (\ref{eq:K-Many}) for our choice of $P$ and $f(R)$ only depends on $X_{Tot}$, using the above expression we finally obtain
\begin{equation}
K(Z_{ij},\phi_k)=\frac{Z_{Tot}(1-\alpha\kappa^2 Z_{Tot})-2V}{1-8\alpha \kappa^2V} \ ,
\end{equation}
which naturally generalizes our previous results to an arbitrary number of scalar fields with arbitrary couplings $X_{ij}$.

For the inverse problem in the quadratic $f(R)$ model we are dealing with we can proceed in exactly the same way as for a single real scalar field. The generalization of Eq.(\ref{eq:GRtoRBG}) for a Lagrangian of the form
\begin{equation}
K(Z_{ij},\phi_k)=Z_{Tot}-2V(\phi_1,\ldots,\phi_N) \ ,
\end{equation}
is now
\begin{equation}\label{eq:Rpal_Nfields}
\frac{R}{f_R}=\kappa^2(Z_{Tot}-4V) \ ,
\end{equation}
and leads to
\begin{equation}
R=\frac{Z_{Tot}-4V}{1-2\alpha(Z_{Tot}-4V)} \ .
\end{equation}
From the relation $X_{Tot}=f_R Z_{Tot}$ with $f_R=1+2\alpha R$ written using the solution of Eq.(\ref{eq:Rpal_Nfields}), one finally finds
\begin{equation}
P(X_{ij},\phi_k)=\frac{X_{Tot}(1+\alpha \kappa^2 X_{Tot})-2V}{1+8\alpha \kappa^2 V} \ ,
\end{equation}
which is in complete agreement with our previous result for a single scalar field in Eq.(\ref{eq:ZGR}).

\subsection{Application 2: EiBI gravity coupled to many scalar fields}

In the EiBI case, the approach slightly differs from the general one. In fact, the relation \eqref{eq:Om-EiBI} specified to the case of many scalar fields with associated energy-momentum tensor \eqref{eq:TmnXMany} provides the following equation:
\begin{equation}
|\hat{\Omega}|^{\frac{1}{2}}{[\Omega^{-1}]^{ \mu}}_\nu = \A{{\delta^\mu}_{\nu}} + \sum_{ij}\B^{ij}X_{ij}{}^\mu{}_\nu \ .\label{eq:OmOm-Many}
\end{equation}
where
\begin{equation}
\A \equiv \lambda +\frac{\epsilon\kappa^2}{2} P\,,\qquad \B^{ij} \equiv - \epsilon\kappa^2 P^{ij}\,.\label{eq:A-B} 
\end{equation}
Taking the determinant of both sides of \eqref{eq:OmOm-Many}, one can find
\begin{eqnarray}
|\hat\Omega| = \A^4+\A^3 \B_1+\B_1^4+\frac12 \A^2\left(\B_1^2-\B_2\right)-\frac14 \B_1^2 \B_2 +\nonumber
\\+ 3\B_2^2 +\frac13\B_1 \B_3 + \frac16 \A\left(\B_1^3-3\B_1 \B_2 + 2 \B_3\right) - \frac14 \B_4 \ ,\nonumber
\\\label{eq:DetOmega}
\end{eqnarray}
where
\begin{eqnarray}
&&\B_1 = \sum_{ij}\B^{ij}X_{ij}\,\quad ;\quad   \B_2 = \sum_{ijkl}\B^{ij}\B^{kl}X_{il}X_{jk}\ ,\nonumber\\
&&\B_3 = \sum_{ijklmn}\B^{ij}\B^{kl}\B^{mn}X_{in}X_{jk}X_{lm}\ ,\label{eq:Bs}\\
&&\B_4 = \sum_{ijklmnpq}\B^{ij}\B^{kl}\B^{mn}\B^{pq}X_{iq}X_{jk}X_{lm}X_{np}\ .\nonumber
\end{eqnarray}

The mapping equation \eqref{eq:K-Many}, specified to the EiBI case, provides the parametrization of the Einstein-frame Lagrangian in terms of RBG quantities as
\begin{equation}\begin{array}{lcl}
&&K[Z_{ij}(X_{mn}),\phi_l]= \label{eq:K-Par-RBT}\qq\\[2mm]
&&\q=\frac{2}{\epsilon \kappa^2}\left\{
1-|\hat\Omega|^{-1/2}\left[\lambda+\frac{\epsilon \kappa^2}{2}\left(P-\sum_{i,j}P^{ij}X_{ij}\right)\right]
\right\}   \end{array}
\end{equation}
which, together with Eq.\eqref{eq:Kij-Many}, allows to find the following explicit form of the functions defined in \eqref{eq:OmOm-Many-Einstein} as
\begin{eqnarray}
\tilde A &=& 1-\frac{\epsilon\kappa^2}{2} \left(K-\sum_{i,j}K^{ij} Z_{ij}\right) \label{eq:AB-GRframe1} \\
\tilde B^{ij}&=&-\epsilon \kappa^2 K^{ij}\,.
\label{eq:AB-GRframe2}
\end{eqnarray}
The scalar sector in the GR-frame is unveiled upon replacing the kinetic term of the scalars $X_{ij}{}^\mu{}_\nu$ from \eqref{eq:Xij-Many} in \eqref{eq:K-Par-RBT}.

Concerning the inverse problem, let us assume that the scalar sector $K(Z_{ij},\phi_l)$ is known. Then, from \eqref{eq:P-Many} one can find the parametrization of the scalar sector in the RBG-frame as
\begin{equation}\begin{array}{lcl}
&&P[X_{ij}(K_{mn},\phi_l),\phi_k] = \qq\\[2mm]
&&\q=\frac{2}{\epsilon \kappa^2}\! \left\{|\hat\Omega|^{1/2} \!\left[ 1 + \frac{\epsilon \kappa^2}{2}\left(\sum_{m,n}K^{mn} Z_{mn} - K\right) \right]\! -\! \lambda\right\}\, 
\end{array}\end{equation}
Finally, using \eqref{eq:Om-Many} where, according to \eqref{eq:OmOm-Many}, one finds $A=|\hat\Omega|^{-1/2}\A$ and $B^{ij}=|\hat\Omega|^{-1/2}\B^{ij}$, it is possible to write the Lagrangian density of the scalar sector in the RBG frame $P(X_{ij},\phi_l)$.

As an example of the above considerations, let us analyse the scalar sector \eqref{eq:canon-2scal} of the previous section. In this case, the determinant of the $\Omega$-matrix \eqref{eq:DetOmega} can be computed in terms of the functions defined in \eqref{eq:Bs} and \eqref{eq:A-B}, that assume the following form
\begin{eqnarray}
\B_1 &=& -\epsilon \kappa^2\left(X_{11}+X_{22}\right) \ , \nonumber\\
\B_2 &=& \left(\epsilon \kappa^2\right)^2\left(X_{11}^2 + 2 X_{12}^2 + X_{22}^2\right)\ ,\\
\B_3 &=& -\left(\epsilon \kappa^2\right)^3\left(X_{11}^3 + 3 X_{11} X_{12}^2 + 3 X_{22} X_{12}^2 + X_{22}^3\right)\ ,\nonumber\\
\B_4 &=& \left(\epsilon \kappa^2\right)^4\left(X_{11}^4 + 4 X_{11}^2 X_{12}^2 +2 X_{12}^4 + 4 X_{11} X_{22} X_{12}^2 +\right.\nonumber\\
&+& \left. 4 X_{22}^2 X_{12}^2 + X_{22}^4\right),\nonumber
\end{eqnarray}
with the definitions
\begin{equation}
\A = \lambda +\frac{\epsilon\kappa^2}{2} \left(X_{11}+X_{22}-2V\right)\,,\quad \B^{11} = \B^{22} = - \epsilon\kappa^2\,,
\end{equation}
that, in turn, completely specify the $\Omega$-matrix in the RBG-frame. It is clear from the above discussion that in the GR-frame, non-linear interaction terms will appear. The relation between the kinetic terms is encoded in \eqref{eq:Zij}, which in matricial notation is given by
\begin{eqnarray}
\left(
  \begin{array}{c}
    Z_{1}{}^\mu{}_\nu \\
    Z_{2}{}^\mu{}_\nu \\
  \end{array}
\right)
=\left(
  \begin{array}{cc}
    {\cal M} & 0 \\
    0 & {\cal M} \\
  \end{array}
\right)
\left(
  \begin{array}{c}
    X_{1}{}^\mu{}_\nu \\
    X_{2}{}^\mu{}_\nu \\
  \end{array}
\right)\ ,
\end{eqnarray}
where we have introduced the matrix
\begin{equation}
{\cal M} = \left(
  \begin{array}{cc}
    A + B X_{11} & B X_{12} \\
    B X_{12} & A + B X_{22} \\
  \end{array}
\right)
\end{equation}
together with the vectors
\begin{equation}
Z_1{}^\mu{}_\nu = \left(
  \begin{array}{c}
    Z_{11}{}^\mu{}_\nu \\
    Z_{21}{}^\mu{}_\nu \\
  \end{array}
\right)\,,\quad
Z_2{}^\mu{}_\nu = \left(
  \begin{array}{c}
    Z_{12}{}^\mu{}_\nu \\
    Z_{22}{}^\mu{}_\nu \\
  \end{array}
\right)\ ,
\end{equation}
and defined the functions
\begin{eqnarray}
&&A \equiv |\hat\Omega|^{-1/2}\A = |\hat\Omega|^{-1/2}\left[\lambda +\frac{\epsilon\kappa^2}{2} \left(X_{11}+X_{22}-2V\right)\right]\,,\nonumber\\
&&B \equiv |\hat\Omega|^{-1/2}\B^{11} = |\hat\Omega|^{-1/2}\B^{22} = - \epsilon\kappa^2 |\hat\Omega|^{-1/2}\,,
\end{eqnarray}
consistently with Eqs. \eqref{eq:Om-Many} and \eqref{eq:OmOm-Many}. Finally, the parametric form of the scalar sector in the GR-frame as given in \eqref{eq:K-Par-RBT}, turns out to be
\begin{eqnarray}
K[Z_{ij}(X_{mn}),\phi_l]=\tfrac{2}{\epsilon \kappa^2}\left\{
1-|\hat\Omega|^{-1/2}\left(\lambda+\epsilon \kappa^2\,V\right)
\right\}.\; \qq
\end{eqnarray}
Due to the dependence on the determinant of the $\Omega$-matrix, one can find $K(Z_{ij},\phi_l)$ just by expressing the $\Omega$-matrix in the GR-frame \eqref{eq:OmOm-Many-Einstein}, using \eqref{eq:AB-GRframe1} and \eqref{eq:AB-GRframe2}. This concludes our analysis of the multi-component scalar field case.

\section{Conclusions and perspectives} \label{sec:IV}

In this work we have investigated a family of metric-affine theories of gravity whose Lagrangian density is a non-linear function of scalars built out of contractions of the metric and the Ricci tensor (Ricci-Based Gravity theories or RBGs),  and coupled to scalar field matter. This family of theories includes many particular cases of interest previously considered in the literature, besides GR itself. It has been shown that the field equations admit an Einstein-frame representation in terms of an auxiliary metric and that, more importantly, the right-hand-side of the resulting Einstein-like equations can be written in the form of a conserved stress-energy tensor in which the scalar matter is coupled to the auxiliary metric, exactly reproducing the structure of Einstein's equation of GR. 

We have made explicit the Einstein-frame representation for the case of a single real scalar field described either by canonical Lagrangians, or by generalized functions of the kinetic and potential terms, finding the general equations of the mapping in those cases. On the gravitational sector, we have chosen two theories of interest, namely, quadratic $f(R)$ theories and Eddington-inspired Born-Infeld gravity. This way, we have explicitly shown how to construct the Einstein-frame matter Lagrangian when the scalar matter model is defined in the original RBG frame. The inverse problem, namely, constructing the matter Lagrangian in the RBG frame if a specific scalar theory is defined in the Einstein frame, has also been worked out in detail. In particular, this has been used to generate specific solutions within a quadratic $f(R)$ model, coupled to a self-gravitating spherically symmetric real scalar field, whose GR counterpart has been known in analytical form since long time ago \cite{Wyman}. The physical properties of such solutions in RBG theories beyond GR will be analyzed in detail in a separate paper. Furthermore, we have extended these results by considering the case of several real scalar fields, where a number of subtleties have been unveiled.


The procedure presented in this work allows to subsequently farming the many astrophysical and cosmological applications of both, canonical and noncanonical scalar-field models in gravitational theories beyond GR. By setting a specific RBG coupled to scalar field matter, and finding the corresponding scalar Lagrangian on the GR side, one can go and solve the GR problem using well established analytical and numerical methods, and transfer the obtained solution to the RBG side via the inverse correspondence. This way, problems and scenarios with scalar fields previously hardly accessible within the context of RBGs by other means, can now be tackled from within GR itself. Moreover, though our analysis has focused on scalar field matter in four spacetime dimensions, it can be extended to other matter sources (for instance, the electromagnetic field case was recently discussed in \cite{Afonso:2018bpv}), and to other dimensions with the necessary adjustments. Furthermore, according to the theorem presented in \cite{Kijowski:2016qbc} following an approach completely different from ours, the extension of this analysis beyond RBGs should also be possible, at least in a formal way. 

The results here presented might be relevant to identify potential degeneracies between exotic matter sources coupled to GR and standard sources coupled to modified theories of gravity, with important implications from both, theoretical and observational perspectives, for instance, within the context of gravitational wave emission. These and other related issues will be explored elsewhere.

\section*{Acknowledgments}

The authors thank Piotr T. Chrusciel for bringing Ref.\cite{Kijowski:2016qbc} to our attention. GJO is funded by the Ramon y Cajal contract RYC-2013-13019 (Spain). DRG is funded by the Funda\c{c}\~ao para a Ci\^encia e a Tecnologia (FCT, Portugal) postdoctoral fellowship No.~SFRH/BPD/102958/2014 and by the FCT research grants No. UID/FIS/04434/2013 and No. PTDC/FIS-OUT/29048/2017.  This work is supported by the Spanish project FIS2014-57387-C3-1-P (MINECO/FEDER, EU), the project H2020-MSCA-RISE-2017 Grant FunFiCO-777740, the project SEJI/2017/042 (Generalitat Valenciana), the Consolider Program CPANPHY-1205388, and the Severo Ochoa grant SEV-2014-0398 (Spain).  This article is based upon work from COST Action CA15117, supported by COST (European Cooperation in Science and Technology).


\begin{thebibliography}{100}

\bibitem{Abbott:2016blz}
  B.~P.~Abbott {\it et al.} [LIGO Scientific and Virgo Collaborations],
  Phys.\ Rev.\ Lett.\  {\bf 116}, 061102 (2016).

  \bibitem{Abbott:2017oio}
  B.~P.~Abbott {\it et al.} [LIGO Scientific and Virgo Collaborations],
  Phys.\ Rev.\ Lett.\  {\bf 119}, 141101 (2017).

  \bibitem{TheLIGOScientific:2017qsa}
  B.~P.~Abbott {\it et al.} [LIGO Scientific and Virgo Collaborations],
  Phys.\ Rev.\ Lett.\  {\bf 119}, 161101 (2017).

  \bibitem{Laureijs:2011gra}
  R.~Laureijs {\it et al.} [EUCLID Collaboration],
  arXiv:1110.3193 [astro-ph.CO].

  \bibitem{Amendola:2016saw}
  L.~Amendola {\it et al.},
  Living Rev.\ Rel.\  {\bf 21}, 2 (2018).


  \bibitem{Barack:2018yly}
  L.~Barack {\it et al.},
  arXiv:1806.05195 [gr-qc].

  \bibitem{SeyYag18}
  B. Seymour and K. Yagi, arXiv:1808.00080 [gr-qc]

  \bibitem{Collett:2018gpf}
 T.~E.~Collett {\it et al.},
  Science {\bf 360},  1342 (2018).

\bibitem{Ishak:2018his}
  M.~Ishak,
  arXiv:1806.10122 [astro-ph.CO].

  \bibitem{Will:2014kxa}
  C.~M.~Will,
  Living Rev.\ Rel.\  {\bf 17}, 4 (2014).

  \bibitem{Lombriser:2015sxa}
  L.~Lombriser and A.~Taylor,
  JCAP {\bf 1603}, 031 (2016).

\bibitem{Lombriser:2016yzn}
  L.~Lombriser and N.~A.~Lima,
  Phys.\ Lett.\ B {\bf 765}, 382 (2017).

 \bibitem{Baker:2017hug}
  T.~Baker, E.~Bellini, P.~G.~Ferreira, M.~Lagos, J.~Noller and I.~Sawicki,
  Phys.\ Rev.\ Lett.\  {\bf 119},  251301 (2017).

\bibitem{Sakstein:2017xjx}
  J.~Sakstein and B.~Jain,
  Phys.\ Rev.\ Lett.\  {\bf 119},  251303 (2017).

  \bibitem{Creminelli:2017sry}
  P.~Creminelli and F.~Vernizzi,
  Phys.\ Rev.\ Lett.\  {\bf 119}, 251302 (2017).

   \bibitem{Ezquiaga:2017ekz}
  J.~M.~Ezquiaga and M.~Zumalac\'arregui,
  Phys.\ Rev.\ Lett.\  {\bf 119}, 251304 (2017).

  \bibitem{EzZuma}
    J.~M.~Ezquiaga and M.~Zumalac\'arregui, arXiv:1807.09241 [astro-ph.CO].

\bibitem{Ferraro:2006jd}
  R.~Ferraro and F.~Fiorini,
  Phys.\ Rev.\ D {\bf 75},  084031 (2007).

 \bibitem{Maluf:2013gaa}
  J.~W.~Maluf,
  Annalen Phys.\  {\bf 525}, 339 (2013).
  
  \bibitem{Bamba:2013ooa}
  K.~Bamba, S.~Capozziello, M.~De Laurentis, S.~Nojiri and D.~S\'aez-G\'omez,
  Phys.\ Lett.\ B {\bf 727},  194 (2013).

  \bibitem{BeltranJimenez:2017tkd}
  J.~Beltr\'an~Jim\'enez, L.~Heisenberg and T.~Koivisto,
  Phys.\ Rev.\ D {\bf 98},  044048 (2018).
  
  \bibitem{BeltranJimenez:2018vdo}
  J.~Beltr\'an~Jim\'enez, L.~Heisenberg and T.~S.~Koivisto,
  JCAP {\bf 1808},   039 (2018).


\bibitem{Jarv:2018bgs}
  L.~Jarv, M.~Runkla, M.~Saal and O.~Vilson,
  Phys.\ Rev.\ D {\bf 97}, 124025 (2018).


\bibitem{Olmo:2011uz}
  G.~J.~Olmo,
  Int.\ J.\ Mod.\ Phys.\ D {\bf 20}, 413 (2011).

\bibitem{Deruelle}
N. Deruelle and L. Farina-Busto, Phys. Rev. D \textbf{41}, 3696 (1990).

\bibitem{Borunda:2008kf}
  M.~Borunda, B.~Janssen and M.~Bastero-Gil,
  JCAP {\bf 0811}, 008 (2008).

\bibitem{Charmousis08}
C. Charmousis, Lect. Notes Phys. \textbf{769}, 299 (2008).

\bibitem{Burton:1997sj}
  H.~Burton and R.~B.~Mann,
  Phys.\ Rev.\ D {\bf 57}, 4754 (1998).

  \bibitem{Odintsov:2014yaa}
  S.~D.~Odintsov, G.~J.~Olmo and D.~Rubiera-Garcia,
  Phys.\ Rev.\ D {\bf 90}, 044003 (2014).


\bibitem{BeltranJimenez:2017uwv}
  J.~Beltran Jimenez, L.~Heisenberg, G.~J.~Olmo and D.~Rubiera-Garcia,
  JCAP {\bf 1710}, 029 (2017).

  \bibitem{Olmo:2011ja}
  G.~J.~Olmo and D.~Rubiera-Garcia,
  Phys.\ Rev.\ D {\bf 84}, 124059 (2011).

  \bibitem{Olmo:2013gqa}
  G.~J.~Olmo, D.~Rubiera-Garcia and H.~Sanchis-Alepuz,
  Eur.\ Phys.\ J.\ C {\bf 74},  2804  (2014).

  \bibitem{Bambi:2015zch}
  C.~Bambi, A.~Cardenas-Avendano, G.~J.~Olmo and D.~Rubiera-Garcia,
  Phys.\ Rev.\ D {\bf 93}, 064016 (2016).
  
  \bibitem{Shaikh:2015oha}
  R.~Shaikh,
  Phys.\ Rev.\ D {\bf 92},  024015 (2015).
  
  \bibitem{Wei:2014dka}
  S.~W.~Wei, K.~Yang and Y.~X.~Liu,
  Eur.\ Phys.\ J.\ C {\bf 75},  253 (2015).
  
  \bibitem{Chen:2018mkf}
  C.~Y.~Chen and P.~Chen,
  Phys.\ Rev.\ D {\bf 98},  044042 (2018).

\bibitem{Afonso:2018bpv}
  V.~I.~Afonso, G.~J.~Olmo and D.~Rubiera-Garcia,
  Phys.\ Rev.\ D {\bf 97}, 021503 (2018).

\bibitem{Afonso:2018mxn}
  V.~I.~Afonso, G.~J.~Olmo, E.~Orazi and D.~Rubiera-Garcia,
  arXiv:1807.06385 [gr-qc].

  \bibitem{Macedo:2013jja}
  C.~F.~B.~Macedo, P.~Pani, V.~Cardoso and L.~C.~B.~Crispino,
  Phys.\ Rev.\ D {\bf 88}, 064046 (2013).

  \bibitem{Brito:2015pxa}
  R.~Brito, V.~Cardoso, C.~A.~R.~Herdeiro and E.~Radu, Phys.\ Lett.\ B {\bf 752}, 291 (2016).

  \bibitem{Herdeiro:2018daq}
  C.~Herdeiro, I.~Perapechka, E.~Radu and Y.~Shnir,  arXiv:1808.05388 [gr-qc].

 \bibitem{Cunha:2018acu}
  P.~V.~P.~Cunha and C.~A.~R.~Herdeiro,  Gen.\ Rel.\ Grav.\  {\bf 50}, 42 (2018).

\bibitem{Herdeiro:2015waa}
  C.~A.~R.~Herdeiro and E.~Radu,
  Int.\ J.\ Mod.\ Phys.\ D {\bf 24},  1542014 (2015).

  \bibitem{ArmendarizPicon:1999rj}
  C.~Armendariz-Picon, T.~Damour and V.~F.~Mukhanov,
  Phys.\ Lett.\ B {\bf 458}, 209 (1999).

  \bibitem{ArmendarizPicon:2000ah}
  C.~Armendariz-Picon, V.~F.~Mukhanov and P.~J.~Steinhardt,
  Phys.\ Rev.\ D {\bf 63}, 103510 (2001).


\bibitem{Bazeia:2007df}
  D.~Bazeia, L.~Losano, R.~Menezes and J.~C.~R.~E.~Oliveira,
  Eur.\ Phys.\ J.\ C {\bf 51}, 953 (2007).

\bibitem{Starobinski80}
A. A. Starobinsky, Phys. Lett. B \textbf{91},  99 (1980).

\bibitem{BanFer10}
M. Banados, P. G. Ferreira, Phys. Rev. Lett. \textbf{105}, 011101 (2010).

\bibitem{Afonso:2017bxr}
  V.~I.~Afonso, C.~Bejarano, J.~Beltran Jimenez, G.~J.~Olmo and E.~Orazi,
  Class.\ Quant.\ Grav.\  {\bf 34}, 235003 (2017).

\bibitem{Barrientos:2018cnx}
  E.~Barrientos, F.~S.~N.~Lobo, S.~Mendoza, G.~J.~Olmo and D.~Rubiera-Garcia,
Phys. Rev. D \textbf{97},  104041 (2018).

\bibitem{AbbottNS}
B. P. Abbott {\it et al.} [LIGO Scientific and Virgo Collaborations], The Astrophysical Journal Letters {\bf 848}, L13 (2017).

\bibitem{Banados}
M. Ba\~nados, P. G. Ferreira, and C. Skordis, Phys. Rev. D \textbf{79}, 063511 (2009).

\bibitem{BI1}
P. Pani, T. Delsate, and V. Cardoso, Phys. Rev. D \textbf{85},  084020 (2012).


\bibitem{BI2}
S.~Jana and S.~Kar,  Phys.\ Rev.\ D {\bf 92},  084004 (2015).

\bibitem{BI4}
C.~Y.~Chen, M.~Bouhmadi-Lopez, and P.~Chen, Eur.\ Phys.\ J.\ C {\bf 76}, 40 (2016).

\bibitem{BI5}
S.~L.~Li and H.~Wei, Phys.\ Rev.\ D {\bf 96}, 023531 (2017).

\bibitem{BI5b}
  M.~Roshan, A.~Kazemi and I.~De Martino,
  Monthly Notices of the Royal Astronomical Society, \textbf{479}, 1287 (2018).

\bibitem{BI6}
  C.~Y.~Chen and P.~Chen, Phys.\ Rev.\ D {\bf 98}, 044042 (2018).

\bibitem{BI7}
  R.~Shaikh, Phys.\ Rev.\ D {\bf 98}, 064033 (2018).

\bibitem{Gu:2018lub}
  B.~M.~Gu, Y.~X.~Liu and Y.~Zhong,  Phys.\ Rev.\ D {\bf 98}, 024027 (2018).

\bibitem{BI8}
  S.~Jana, R.~Shaikh and S.~Sarkar, arXiv:1808.09656 [gr-qc].

\bibitem{Bouhmadi-Lopez:2018sto}
  M.~Bouhmadi-Lopez, C.~Y.~Chen, P.~Chen and D.~h.~Yeom,  arXiv:1809.06579 [gr-qc].


\bibitem{BeltranJimenez:2017doy}
  J.~Beltran Jimenez, L.~Heisenberg, G.~J.~Olmo and D.~Rubiera-Garcia,
  Phys.\ Rept.\  {\bf 727}, 1 (2018).

  \bibitem{BI5c}
S.~Jana, G.~K.~Chakravarty and S.~Mohanty,  Phys.\ Rev.\ D {\bf 97},  084011 (2018).

\bibitem{Afonso:2017aci}
  V.~I.~Afonso, G.~J.~Olmo and D.~Rubiera-Garcia, JCAP {\bf 1708}, 031 (2017).


  \bibitem{BI34}
M. Born, L. Infeld, Proc. Roy. Soc. London. A \textbf{144}, 425 (1934).

\bibitem{Felder:2002sv}
  G.~N.~Felder, L.~Kofman and A.~Starobinsky,
  JHEP {\bf 0209}, 026 (2002).

\bibitem{Jana:2016uvq}
  S.~Jana and S.~Kar,
  Phys.\ Rev.\ D {\bf 94},  064016 (2016).

\bibitem{Wyman}
M. Wyman, Phys. Rev. D \textbf{24}, 839 (1981).


\bibitem{Kijowski:2016qbc}
  J.~Kijowski,
  Int.\ J.\ Geom.\ Meth.\ Mod.\ Phys.\  {\bf 13}, 1640008 (2016).


  \end{thebibliography}
\end{document}